\documentclass[a4paper,twocolumn,11pt,accepted=2024-11-14]{./quantumarticle}
\pdfoutput=1

\usepackage[utf8]{inputenc}
\usepackage[english]{babel}
\usepackage[T1]{fontenc}
\usepackage{amsmath}
\usepackage{amssymb}
\usepackage{graphicx,xcolor,colortbl}

\usepackage{mathtools}
\usepackage{xspace}

\newcommand{\psw}{\mathit{SWAP}}
\newcommand{\cz}{\mathit{CZ}}
\newcommand{\crot}{\mathit{CRot}}
\newcommand{\cph}[1]{\mathit{CPh}(#1)}
\newcommand{\zx}{\mathit{ZX}}
\newcommand{\zz}{\mathit{ZZ}}

\usepackage[caption=false]{subfig}
\usepackage{xcolor,colortbl}

\usepackage{tikz}
\usetikzlibrary{quantikz}
\usepackage[notrig]{physics}
\usepackage{bm}

\renewcommand{\vec}[1]{\boldsymbol{\mathrm{#1}}}

\usepackage{booktabs}

\usepackage{hyperref} 
\usepackage[nameinlink,noabbrev]{cleveref}

\usepackage[numbers,sort&compress]{natbib}
\usepackage{array,multirow}
\newcolumntype{C}[1]{>{\centering\arraybackslash}m{#1}}
\newcolumntype{M}[1]{>{\centering\arraybackslash}m{#1}}

\newcommand{\ie}{i.e.,~}
\newcommand{\eg}{e.g.,~}

\newcommand{\cs}{${}^{13}\text{C}$~}
\newcommand{\ns}{${}^{14}\text{N}$~}
\newcommand{\nv}{$\text{NV}^-$~}

\usepackage{adjustbox}
\usepackage{verbatim}

\begin{document}

\title{The Virtual Quantum Device (VQD): A tool for detailed emulation of quantum computers}

\author{Cica Gustiani}
\affiliation{Department of Materials, University of Oxford, Parks Road, Oxford OX1 3PH, United Kingdom}
\orcid{0000-0003-0558-4685}
\email{cicagustiani@gmail.com}

\author{Tyson Jones}
\orcid{0000-0002-9360-5417}

\author{Simon C. Benjamin}
\orcid{0000-0002-7766-5348}
\email{simon.benjamin@materials.ox.ac.uk}
\affiliation{Department of Materials, University of Oxford, Parks Road, Oxford OX1 3PH, United Kingdom}
\affiliation{Quantum Motion, 9 Sterling Way, London N7 9HJ, United Kingdom}

\begin{abstract}
We present the Virtual Quantum Device (VQD) platform, a system based on the QuEST quantum emulator. Through the use of VQDs, non-expert users can emulate specific quantum computers with detailed error models, bespoke gate sets and connectivities. The platform boasts an intuitive interface, powerful visualisation, and compatibility with high-performance computation for effective testing and optimisation of complex quantum algorithms or ideas across a range of quantum computing hardware. We create and explore five families of VQDs corresponding to trapped ions, nitrogen-vacancy-centres, neutral atom arrays, silicon quantum dot spins, and superconducting devices. Each is highly configurable through a set of tailored parameters.  We showcase the key characteristics of each virtual device, providing practical examples of the tool's usefulness and highlighting each device's specific attributes. By offering user-friendly encapsulated descriptions of diverse quantum hardware, the VQD platform offers researchers the ability to rapidly explore algorithms and protocols in a realistic setting; meanwhile hardware experts can create their own VQDs to compare with their experiments.
\end{abstract}

\section{Introduction}

In this paper we introduce the Virtual Quantum Device (VQD) platform, a system for emulating noisy quantum computers.
The VQD supports realistic error
models and bespoke operations with the flexibility to tailor to the diverse range of quantum hardware currently being investigated worldwide. Alongside the VQD platform itself, we have worked with collaborators to establish the following VQD instances: multi-node ion traps, nitrogen-vacancy-centre (NV-centre) diamond qubits, neutral atoms, silicon qubits, and superconducting qubits.  

The VQD platform is built atop
{QuESTlink}~\cite{jones2020questlink}, a {Mathematica}
extension of {QuEST}~\cite{jones2019quest}, itself an open-source emulator of quantum computers developed in \texttt{C} and \texttt{C++}. 
A user can access these tools through a Mathematica notebook with the appropriate licence, or through the Wolfram Engine which is presently free for developers.
{QuESTlink} combines {Mathematica}'s
powerful symbolic operations with {QuEST}'s high-performance backend, enabling virtual devices to be highly configurable through an intuitive interface, and able to leverage powerful visualisation facilities, without compromising performance.
Through the VQD platform, researchers can effectively test and optimise complex quantum
algorithms or ideas across a range of quantum computing architectures.
In its current form, the {QuESTlink} tool's capability to exactly model a system of qubits is limited only by the host computer's memory. For example a 64\,GiB RAM system can emulate up to 32 qubits in double precision, or about half as many if one uses the full density matrix representation rather than state vectors~\footnote{In future, {QuESTlink} may be deployed in distributed settings, as already supported by the underlying {QuEST} simulator.}. As we presently explain, the VQD platform is flexible with respect to this choice.

In the VQD platform, end-users can easily select the particular, pre-existing VQD that they wish to emulate (a given form of ion trap, for example). Users can re-configure an instance of a virtual
device by adjusting a set of parameters that are tailored to each type of
device; some of these parameters will be immediately comprehensible to a non-expert user. Other parameters are provided for the more expert users to adjust if they wish.
The parameters are carefully chosen to capture the main
characteristics and describe the critical error sources of the device. For
instance, the parameters may include the number of qubits and nodes, atom
locations, relaxation times $T_1$ and $T_2$, qubit frequencies, gate
fidelities, Rabi frequencies, and more. Ultimately of course a user can create their own VQD as a variant of one of the existing devices or as a wholly new class of system. Thus, for example, a research team who have developed a specific type of device might do this in order to allow others to experiment on their system {\it virtually} and find interesting tasks for it -- or simply as an internal tool to closely model their experiments.

To ensure an accurate and realistic simulation, the virtual device is designed
to reflect a close approximation to the physical reality by only providing
access to physically possible native operations.  For example, in the case of
neutral atoms, multi-qubit gates can only be implemented when the atoms are within the overlapping blockade radii. 
The corresponding VQD tracks the atom's spatial configurations and movements, to ensure that the operations called for by a user are physically possible.  

Typically the user will input a quantum circuit composed of legitimate operations, and the tool can output the noise-decorated circuits according to the instantiated virtual device, as well as the final state of the virtual
device, such as the final arrangement of atoms in neutral atom architectures.

This paper presents several demonstrations utilising the VQD platform which highlight the unique features of each qubit implementation. These demonstrations serve as practical examples of the tool's usefulness and exhibit the specific
attributes of each device. For instance, the demonstration of entanglement distillation on trapped ions showcases the long coherence times and precise laser control capabilities of these qubits. The simulation of BSC theory on
diamond NV-centers emphasises the central spin system connectivity of these
qubits. The demonstration of graph state preparation on neutral atoms
highlights the mobility and two-dimensional connectivity of these qubits. By reproducing the published experiment results involving Bell state preparation on silicon
qubits, we demonstrate that our error model accurately captures the device.
Lastly, the variational algorithm demonstration on superconducting qubits
highlights the potential for these qubits to implement complex algorithms as well as the importance of tailoring circuits to the system's connectivity.

The VQD code is readily accessible for use here~\cite{vqd}. Currently in its inaugural beta stage, the code is fully functional and can be employed to reproduce the results presented in this paper. It is important to note that while this version is operational, it is not the final form. Future versions will reflect iterative enhancements and refinements. This presents a unique opportunity for users to engage with the code in its developing stages while being capable of effectively using the functionalities presented here.

\section{The virtual quantum device package}

    A virtual quantum device provides a layer of abstraction above QuESTlink's \textit{device specification} facility introduced in version 0.6~\cite{jones2020questlink}. The latter is a static description of how a realistic, error-prone and constrained quantum device might realise an intended unitary circuit; through a proximate channel composed of mixing operators and a potentially distinct set of unitary operators. For instance, consider how a simple 3-qubit 3-gate circuit might be affected on a noisy device.
    \begin{center}
    \includegraphics[width=\columnwidth]{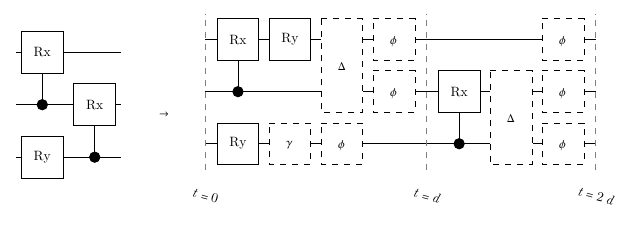}
    \end{center}
    Specifically, the device imperfections in this example are such that gates on the bottom qubit induce damping (signified by $\gamma$), those on the bottom and middle induce depolarising ($\Delta$), and upon the top two induce depolarising and an erroneous rotation about the Y axis. All idle qubits experience dephasing noise ($\phi$) while waiting for the slowest operation in their automatically scheduled parallel batch (in the first round; with duration $d$). Upon a given input state, the right-hand-side channel may be simulated through the density matrix formalism to completely determine the \textit{noisy}, the real-world output of the ideal left circuit.

    In this way, a device specification offers a mapping between ideal unitary circuits and realistic channels admitted by the modelled hardware device.
    Device specifications include provisions of active and passive noise behaviour, restricted gate sets and qubit topologies, gate substitutions, durations and scheduling, time-dependent noise,  general variable-dependent noise (such that noise severity may depend on preceding operations) and \textit{hidden qubits} to model dynamics outside the computational state space. Given a collection of pre-made specifications, they permit a user to rapidly study the differing behaviours of realistic quantum computers when tasked with executing the same circuit.

{
    While simple and powerful to \textit{users} of a device specification, \textit{producing} a specification (as per the guide here~\cite{vqdguide}) 
    has several potential pain points. In \Cref{appendix:designPhilosophyAndQL} we describe these challenges, and how they motivated the choice of QuESTlink as the platform upon which to build the VQD framework. There we explain that QuESTlink's unique combination of a well-featured emulator (QuEST) with Mathematica as a powerful -- albeit commercial -- symbolic manipulation tool, provided elegant design solutions to some rather complex challenges. 
}
    
   A particular feature of QuESTlink which we stress here is its capability to switch between state vector and density matrix representations. The latter can comprehensively represent the impact of noise processes on a quantum state, and QuESTlink supports general Kraus operators acting on the density matrix to fully encode those processes.   However the representation is 
  quadratically more expensive than statevector simulation of the original noise-free circuit, and this is sometimes prohibitive.
    Fortunately, these steeper memory and time costs can often be avoided by the user instead opting for Monte Carlo statevector simulation, whereby expected values of the VQD prescribed noisy quantum states are evaluated through repeated sampling of an ensemble of error-affected statevectors~\cite{grurl2022noise}. 
    QuESTlink's \texttt{SampleExpecPauliString} function pseudo-randomly replaces each decoherence channel with an operator from its Kraus map, and acts it upon a statevector before computing the expected value. This process is repeated until the average of the samples converges. Thus, while virtual quantum devices are typically specified in terms of operations that are incompatible with the state vector representation, and indeed the full density matrix representation is available, a user can opt to {\it automatically translate} the emulation task into the Monte Carlo paradigm. This often enables simulation of realistic noisy devices at modest memory costs.

\section{Error models}

In this section, we elaborate the error models and forms of noise considered in our virtual devices. Each noise model can be described using Kraus maps and classical descriptions. Kraus maps mathematically describe the evolution of quantum states resulting from interactions with the environment, while classical descriptions incorporate errors that can be observed in real-life situations. For instance, in neutral atoms, the atoms can be completely ejected into the environment. Such a situation can be straightforwardly indicated by suitable flags.

By combining these approaches, we provide a tool with a sophisticated error
model that is powerful but with a friendly user interface. This enables us to
easily simulate and analyse various scenarios or quantum algorithms on virtual
qubits, with accurate descriptions of the effects of noise and other error
sources.  It is important to note that our models do include some basic simplifications with respect to what occurs in real experiments; we have made each of our demonstration VQDs sufficiently complex to capture much of the uniqueness of each platform, but there are many respects in which they can be made more sophisticated by a user who is expert in the relevant system.  In the following, we outline the simplifying assumptions
that apply to our demonstration virtual devices.

Firstly, {
each operator acting on a qubit is subjected to the rotating-wave approximation implemented within its own rotating frame of reference; }
thus, in the absence of any gate application or
noise, the qubit simply undergoes the identity operation. Secondly, when the
user provides the $T_2$ time, our default assumption is that any required Hahn-echo or dynamical decoupling is understood to be
applied `in the background' to the passive qubits -- that is to say, such decompiling is not explicitly generated in the circuits produced by the VQD platform. Lastly, it is important to recognise that the QuEST family of tools is designed to represent only discrete events occurring to a series of qubits -- thus other tools must be used to, \eg model the continuous-time dynamics of an $n$-level system subject to given Hamiltonian pulse. (However, having understood the effect of that pulse within a pair of levels deemed to constitute a qubit, then the resulting unitary or non-unitary operation can certainly be described within a VQD.)

A final remark is that the density matrices
presented in this paper (\eg \Cref{fig:dist_best,fig:Bells}) are visualisations generated directly from the density matrix representation in the VQD tool and are not modified to account for any imperfections in experimental tomography.

\subsection{Standard forms of noise\label{sec:standard_forms}}

We begin by describing textbook standard forms of noise for one- and two-qubit gates.
They include one- and two-qubit depolarising and dephasing and one-qubit amplitude damping. While the VQD platform supports the use of arbitrary Kraus maps to represent general non-unitary processes, these three particular forms of noise are so common that it is useful to overtly specify their implementation within the VQD (and indeed, there are optimised algorithms for implementing these specific forms of noise within the underlying QuEST tools). 

Depolarising and dephasing channels are typically attributed to gate
noise. Amplitude damping channel is often attributed to noise in initialisations and
measurement processes, or to continuous relaxation of passive qubits. These forms of noise are straightforward to implement
yet powerful in describing various error processes happening in the experiments. Note that the quantum channels below can be viewed as completely
positive trace-preserving (CPTP) maps.

In the following, we denote $\rho$ as a quantum state, $p$ as the error parameter,
and $\sigma\in\{X,Y,Z\}$ are Pauli operators. The \emph{one-qubit depolarising channel} on qubit $j$ ($depol_j$)
maps
\begin{equation}\label{eq:depol1}
    \rho \mapsto (1-p)\rho + \frac{p}{3}\sum_\sigma{\sigma_j\rho\sigma_j},
\end{equation}
where $p\in [0,\frac{3}{4}]$. 
The \emph{two-qubit depolarising channel} acts homogeneously on qubits $i$ and $j$ ($depol_{i,j}$), and  
maps
\begin{equation}\label{eq:depol2}
\rho \mapsto  (1-p)\rho + \frac{p}{15}(
\sum_\sigma{\sigma_i\rho\sigma_i}+{\sigma_j\rho\sigma_j}
+\sum_{\sigma\sigma'}{\sigma_i\sigma'_j\rho\sigma_i\sigma'_j}),
\end{equation}
where $p\in [0,\frac{15}{16}]$. The \emph{one-qubit dephasing channel}
acts on qubit $j$ ($deph_j$) and maps 
\begin{equation}\label{eq:deph1}
\rho \mapsto (1-p)\rho + p Z_j\rho Z_j,
\end{equation}
where $p\in [0,\frac{1}{2}]$. The \emph{two-qubit dephasing 
channel} acts homogeneously on qubits $i$ and $j$ ($deph_{i,j}$) and maps 
\begin{equation}\label{eq:deph2}
\rho \mapsto (1-p)\rho + \frac{p}{3}(
    Z_i\rho Z_i+
    Z_j\rho Z_j+
    Z_iZ_j\rho Z_i Z_j
    ),
\end{equation}
where $p\in [0,\frac{3}{4}]$.

The \emph{amplitude damping channel} on qubit $j$ ($amp_j$) corresponds to the error map 
\begin{equation}
    \label{eq:damp}
    \rho\mapsto\sum_{A}A_j\rho A_j^\dagger,
A\in
\left\{\begin{pmatrix}
        1 & 0 \\
        0 & \sqrt{1-p}
    \end{pmatrix}
    ,
    \begin{pmatrix}
        0 & \sqrt{p} \\
        0 & 0
\end{pmatrix}\right\},
\end{equation}
where $p\in[0,1]$; for the case $p=1$, the quantum state $\rho$ will be in the state $\ketbra0$. 

\subsection{Standard error parameter estimate from average gate fidelity}\label{sec:standard_error}

In configuring our virtual quantum devices, we often encounter the situation that the experimental team's best model for the noise in a given process is of the depolarising and/or dephasing type.
We deem an operation to have \emph{standard error} when its noise model can be described by the composition  
\begin{equation}\label{eq:stderr}
    \mathcal E(p,q) \equiv deph(p)\circ depol(q) 
\end{equation}
where $depol(p)$ and $deph(q)$ are one- or two-qubit parameterised depolarising and dephasing noise 
defined in \Cref{sec:standard_forms}).
It is then very valuable to be able to take fidelity estimates obtained experimentally and translate them into the noise severities $p$ and $q$. Fortunately, this is straightforward, and the procedure is described in Appendix \ref{appendix:depolDephase}.

\subsection{Off-resonant Rabi oscillation}

Many quantum gate operations are realised by applying a coherent pulse of EM (\eg microwave) radiation at a frequency resonant with the energy gap of the qubit. Ideally, the qubit is perfectly isolated during the process in order to prevent the unwanted driving of other qubits. However, as this ideal is generally not achieved, we define the error induced
by off-resonant frequency on the idle qubits during the drive. 

Let $q$ be the actively driven qubit with Rabi frequency $\Omega$ for a duration
$t$. The passive qubit $j$ with frequency difference $\Delta$ receives an off-resonant 
drive described with the following unitary matrix  
\begin{widetext}
\begin{equation}\label{eq:off_resonant}
    \begin{bmatrix}
        \left(\cos(\frac{\Omega_Rt}{2}) - \frac{i\Delta}{\Omega_R}\sin(\frac{\Omega_Rt}{2})\right)e^{i\Omega_R t/2}
        &
        -\frac{i\Omega}{\Omega_R}\sin(\frac{\Omega_Rt}{2})e^{i\Omega_R t/2}
        \\
        -\frac{i\Omega}{\Omega_R}\sin(\frac{\Omega_Rt}{2})e^{-i\Omega_R t/2}
        &
        \left(\cos(\frac{\Omega_Rt}{2}) + \frac{i\Delta}{\Omega_R}\sin(\frac{\Omega_Rt}{2})\right)e^{-i\Omega_R t/2}
\end{bmatrix},
\end{equation}
\end{widetext}

where $\Omega_R=\sqrt{\Omega^2+\Delta^2}$ {
is the detuned frequency. \Cref{eq:off_resonant} describes Rabi oscillation of two-level system (states $\ket0$ and $\ket1$) in the frame of qubit $q$, such that 
it approaches identity for a large} frequency difference $\Delta$. It is thus desirable 
for qubits to have significantly diverse Rabi frequencies.

\subsection{Crosstalk}

An ideal quantum computer would localise gate operations to only the qubits which are nominally involved, and only for the defined duration of the gate. In reality, this localisation may be imperfect either in time (leading to a residual interaction even in the `off' state) or in space whereby `bystander' qubits are brought into the process. Generally, these effects are addressed as \emph{crosstalk}. Two forms of crosstalk occur in the VQDs that we discuss presently.

The first type is in the form of \emph{conditional phase} on bystander qubits, according to the state of a given qubit. This effect may occur only during a gate operation, or it may be a continuous process constituting a failure to decouple the qubits entirely.
In either case, we express this
error between qubits $p$ and $q$, in the form of conditional $Z$-rotation,
\begin{equation}\label{eq:weak_dephasing}
    \ketbra0_p\otimes I_q +\ketbra 1_p\otimes Rz(\alpha)_q,
\end{equation}
or the $ZZ$-coupling, 
\begin{equation}\label{eq:weak_dephasing_zz}
    e^{-i \alpha Z_p Z_q},
\end{equation}
where $\alpha$ is the error parameter which may be dependent on the duration of any active gates, the spatial location of bystanders, etc.

The second type is \emph{unwanted exchange interaction}, a commonly encountered crosstalk model that occurs during the operation of a two-qubit gate and results in partial coupling with bystanders. It has the form of 
\begin{equation}\label{eq:exchange_pauli}
    e^{-i(X_iX_j+Y_iY_j+Z_iZ_j)\theta/4},
\end{equation}
where the error parameter $\theta$ depends on the duration of 
the operated two-qubit gate.

 {There are of course many other cross-talk effects, each dependent on the physics of the system. For example, unwanted partial entanglement with bystander qubits, as may occur in laser-driven two qubit ion trap gates~\cite{herold2016universal,wu2018noise}, has elements in common with the two phenomena described above but is distinct from either. The VQDs we introduce can be extended to accommodate any such effects, by extending and adapting the established examples.}

\subsection{Passive noise and free induction decay}

Within all our VQDs we include a process of \emph{passive noise} \ie noise operations to which qubits are subjected while they are not actively part of an operation; typically the passive noise has a time parameter corresponding to the duration of the active procedure. Passive noise can have any form within the VQD framework, and can be effectively `overwritten' by processes such as crosstalk that engage the nominally passive qubits with the active ones. However, in several cases the passive noise is of a simple form that captures an isolated qubit's decay processes. 

The coherence time of an individual qubit is commonly characterised by the loss of
coherence in the $xy$-axis (transversal relaxation) and $z$-axis (longitudinal
relaxation).  In practice, these characteristics are commonly identified by how
long a quantum state can maintain the $\ket{+}$ state for the traversal
relaxation and the $\ket{1}$ state for the longitudinal relaxation.  

In many systems the more rapid process is the traversal relaxation characterised by $T_2$ (or $T_2^*$ when there is no mitigating procedure, such as Hahn echo, employed). This decay describes the
loss of phase information and ultimately renders a state into a classical mixture of its basis components. In our VQDs we employ to model each with dynamic error parameters
growing with time $\Delta t$ as follows 
\begin{align}
    \mathit{deph}\left(\frac{1}{2}(1-e^{-\Delta t/T_2})\right) \label{eq:depht2}    \\
    \mathit{deph}\left(\frac{1}{2}(1-e^{-(\Delta t/T_2^*)^2})\right) \label{eq:depht2s}
\end{align}
for the exponential $T_2$ and the gaussian $T_2^*$ decays, respectively.
In practice, the $T_2^*$ is much shorter than the $T_2$. The $T_2^*$ characteristic can be ``echoed-out''
to $T_2$ by continuously applying Hahn echo pulses to passive qubits; this technique is also
known as \emph{dynamical decoupling}. In our models, we typically assume that any such process is applied `in the background' (and imperfections in that process are captured by suitably adjusting the passive noise operations) so that 
 $T_2$ is the relevant timescale. Nevertheless, certain VQDs have the option of disabling this assumption and thus experiencing the `raw' $T_2^*$ process.

In the majority of qubit platforms, the longitudinal relaxation time $T_1$ is significantly longer than the $T_2$ (superconducting qubits being an exception).  Depending on the typically observed events,
the $T_1$ decay is approximated with depolarising noise or amplitude damping with error parameters
increasing with duration $\Delta t$: \begin{equation}\label{eq:depolt1}
\mathit{depol}\left(\frac{3}{4}(1-e^{-\Delta t/T_1}) \right) \end{equation}
\begin{equation}\label{eq:ampt1}
\mathit{amp}\left(1-e^{-\Delta t/T_1}\right)
\end{equation}
for depolarising and amplitude damping models, respectively.   The amplitude
damping model is used on systems where the qubits decay to a lower energy state (that defines $\ket 0$) after the $T_1$ time.  The depolarising error is used when the qubits are completely mixed after the $T_1$ time.

\subsection{SPAM errors}\label{sec:spam}

State preparation and measurement (SPAM) involve similar processes for many
systems. In many cases, the same error characteristic constitutes state preparation
and qubit readout processes. Typical SPAM error models involve
amplitude damping and bit-flip errors. 

The amplitude damping (\Cref{eq:damp}) characterises asymmetric error that favours one of the states. For instance, readouts that involve detecting
photons, in which photon loss is more probable than dark counting. 

On the other hand, bit-flip noise captures error that happens symmetrically,
which is described with random $X$ application 
\begin{equation}\label{eq:bf}
    \rho\mapsto(1-b)\rho+b X \rho X,
\end{equation} 
where $b\in[0,0.5]$ is the probability of the error to occur -- notice that $b=0.5$ is the most severe noise, since $b=1$ would correspond to a definite, and therefore perfectly correctable, $X$ operation. 
On two-qubit measurements such as parity measurement on a pair of silicon spin qubits, bit-flip error may be equally probable to happen to either or both qubits. The two-qubit
bit-flip noise on qubits $i,j$ has the following map 
\begin{equation}\label{eq:bf2}
    \rho\mapsto(1-b)\rho+\frac{b}{3} (X_i \rho X_i+X_j\rho X_j+X_iX_j\rho X_iX_j),
\end{equation} 
where $b\in[0,0.75]$ is the meaningful range of probability for an error to occur. 

We emphasise that the above models are merely examples that occur so frequently that it is useful to name and define them; the VQD platform supports completely arbitrary forms of passive noise. Indeed one can opt to set passive noise to zero, and instead include in the definition of each active operation a complete description of the implications for all qubits.

\subsection{Leakage and Loss errors}

In many systems, the two levels deemed to be the qubit $\ket{0}$ and $\ket{1}$ are merely selected states within a large set of physical states of the physical entity representing the qubit. Restricting the entity to the $\{\ket{0} , \ket{1}\}$ subspace may be challenging and imperfectly achieved. The resulting ``leakage errors'' errors correspond to the qubits escaping to a higher level. Alternatively, a ``loss error'' corresponds to the qubit-representing entity even physically disappearing -- for example, a photonic qubit being absorbed within a fibre, or an atom escaping from its trap.

One can consider three types of scenarios. Firstly, when the qubit is partially
leaked to the environment and thus cannot be recovered. This error is described
with a non-trace preserving map. For instance, implementing multi-qubit gates in neutral atoms involves exciting the corresponding atom(s) to a high-level Rydberg state followed by relaxation to the ground state with a high probability. Since the ground state encodes a qubit state, other relaxation scenarios imply that the qubit remains outside the computational basis. Secondly, when the qubit is physically ejected
from the system. This error is captured with a classical parameter --- whether
the qubit is present or absent. Lastly, a leakage that commonly occurs in
solid-state qubits, where it leaks to a higher level and can potentially be
recovered. However, in the VQD specifications discussed here, this event is not recovered during execution -- such a recovery could be introduced in a more advanced VQD variant. The details of each model are discussed in the corresponding device
description.

\section{Trapped ions}\label{sec:trapped_ions}
\subsection{Trapped ions physical system}

The first VQD we discuss is configured to describe a certain type of ion trap system -- namely, one in which two ion traps are bridged by a photonic link to generate entangled pairs.

Linking quantum computations across a quantum network is a desirable capability, whether in order to connect remote systems or simply as a means to scale a quantum computer on a single site. The trapped ion qubit system~\cite{cirac1995quantum} is
considered to be one of the most promising options. This is due to its
potential for
scalability~\cite{kielpinski2002architecture,friis2018observation},
high-fidelity
operations~\cite{harty2014high,PhysRevLett117060504,PhysRevA84030303}, and the
capability to entangle with photons~\cite{krutyanskiy2022entanglement}, making it a leading contender for distributed quantum computing.

Practically all multi-qubit ion trap systems are variants on a common paradigm, \ie a string of ions confined in 
an oscillating quadrupolar electric field, the Paul trap~\cite{paul1990electromagnetic}.
There are however many significant variants 
depending on numerous factors and not least, the specific species of ion (such as calcium, ytterbium, beryllium, magnesium etc).
The ion trap VQD that we now describe should be readily adaptable to different classes of ion traps and does not correspond exactly to any specific experimental system described in the literature. Our intent is to capture relevant operational characteristics for a multi-node (i.e. more than one ion trap) system. Gate operations between ions in different traps can be performed by exploiting, at each trap, an ion species suited for remote entanglement (\eg strontium as in Ref.~\cite{PhysRevLett.124.110501}) and other ions that are well-suited for high fidelity gates and storage (see \eg \cite{Drmota_2023} for relevant inter-species entanglement transfer). Having configured our VQD we will explore the prospects for entanglement purification, a key enabler for distributed quantum computing. We must of course make specific choices for the available operations, noise models and so on, which we now describe.

The preparation of the state in the trapped ion qubit system is typically achieved via optical pumping. The ions are laser-cooled close to the motional ground state, and then selectively excited and allowed to decay until 
the population has decayed with a high probability to the target electronic state, which is not coupled to the excitation laser.  Readout of the qubits is performed via excitation resonant with state $\ket1$ and the detection of the resultant spontaneous photon emission. Dark readout indicates ``0'' and bright readout indicates ``1''. Single and two-qubit gates can be laser or microwave (or RF) driven; the laser-driven method is the reference of our model. 
We employ the full controlled $\pi$-phase rotation ($\cz$), a native two-qubit gate in this system typically realised via the interaction of a far-detuned bichromatic laser field. This field induces spatially and temporally periodic AC-Stark shifts, producing a qubit-state-dependent force that selectively excites and de-excites the motion of the ions within the trap, imparting a geometric phase conditioned on the collective state of the qubits. This is one of the commonly realised forms of two-qubit gate; other gates such as the M{\o}lmer–S{\o}rensen gate\,\cite{PhysRevLett.82.1971} could of course be added to the VQD as a variation.

For the purposes of incorporation into our VQD, we consider remote entanglement as a type of initialisation, where the state of the corresponding ions are erased and set into the Bell state
$\ket*{\Psi^+}=(\ket{01}+\ket{10})/\sqrt{2}$ -- accompanied by noise defined in the error model. The work described in Ref.~\cite{PhysRevLett.124.110501} is an example; there are many variants several of which derive from the heralded entanglement scheme by~\citeauthor{kokbarett2004efficient}\cite{kokbarett2004efficient}.

In our trapped ion system, the trap region is divided into two partitions for \emph{storage} and \emph{entanglement} purposes. The storage regions are used for operations that require memory, such as initialisation, readout, operating gates, and physical swaps. On the other hand, the entanglement region is solely dedicated to performing remote entanglement with another trap through a heralded photonic scheme. As a physical example, the traps mentioned in references~\cite{barrett2004deterministic,huber2008transport}are segmented or partitioned in a similar manner.
Each region is comprised of one or more zones, each physically defined by a number of electrodes and pathways where control signals are delivered to trap and manipulate the ions. The ions can be moved within and between zones through the use of appropriate operations, including \emph{shuttling}, \emph{separation},  \emph{recombination}, and physical SWAP. Shuttling is a linear movement of the ions, separation involves splitting a string of two or more ions into different zones, and recombination is the inverse procedure, necessary before applying a two-qubit gate or swapping the locations of the ions.

Our VQD thus employs a relatively simple shuttling scenario, appropriate to devices that aim at optical linkage for connectivity.
More complex shuttling capabilities have been demonstrated through the application of Quantum Charge-Coupled Devices (QCCD), as evidenced by research conducted within both academic and industrial sectors
~\cite{mosesrace23,malinowski23how,pino2021demonstration}. Penning traps have also recently gained attention as a possible route to 2D shuttling~\cite{jain2023unit}. Systems such as these are interesting targets for VQD simulation in future projects, given the connectivity enabled by such comprehensive shuttling and the corresponding importance of shuttling noise.

\subsection{Trapped ions architecture and native operations}

\begin{figure}[bht]
\centering
    \includegraphics[width=0.95\columnwidth]{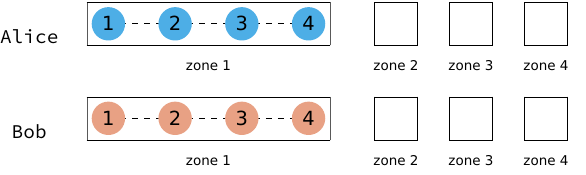}
\caption{
    Two-nodes `Alice' and `Bob', each an ion trap, constituting a single VQD system. Alice and
    Bob each has four ions.
    This figure shows the arrangement of
    the ions when instantiated, where all ions lie in zone 1 and are combined. Combined ions are indicated with dashed
    lines. Qubits initialisation, readout, and physical swaps can be done in
    zones 1, 2, and 3. Quantum gates (single-qubit rotations and two-qubit controlled phase) are
    performed in zones 2 and 3. Remote entanglement is performed in zone 4.  Linear
    shuttles are allowed between any neighbouring zones, but separation and recombination cannot be performed in zone 4.
}
\label{fig:tions}
\end{figure}

We model the trapped ions as a multi-node system, where the user may specify
the nodes and the number of ions in each node. For instance, \Cref{fig:tions}
shows a two-node trap we use in our entanglement distillation simulation. The
traps are identical and comprise four zones with the same assignments in every
segment.

Qubit initialisation to fiducial state $\ket0$ can be done 
to all ions in a single zone at any time, for zones 1, 2 and 3. Projective measurement (in the computational basis) can also be done at any time to any ion in zones 1, 2, and 3; however,
the ion must be alone in the zone during the readout, as we assume that no individual-ion-resolved fluorescence detection is available. Remote entanglement
initialisation to the Bell state $\ket*{\Psi^+}$ is done between two parties,
and the ions must be in zone 4.

Elementary single-qubit operations include single rotations $\{Rx_j(\theta),
Ry_j(\theta), Rz_j(\theta)\}$, where the $x$- and $y$- rotations can be
performed when the ion $j$ lies in zone 2 or 3. The $z$-rotations are implemented virtually as in Ref.~\cite{mckay2017efficient}, which are accounted for in the phase of the subsequent gates; thus, $z$-rotation gates are perfect and instantaneous.

The two-qubit native operations on qubits $i$ and $j$ include the
controlled-phase gate $\cz_{i,j}$ and physical swap $\psw_{i,j}$. To perform
these operations, ions $i$ and $j$ must be combined beforehand.  However, the
implementation of $\cz$ and $\psw$ gates is different. $\cz$ gates can only be
performed in zones 2 and 3, while physical $\psw$ can be performed in zones 1,
2, and 3.

Ions shuttling is a linear movement, meaning that ions can be moved to another
zone as long as there are no other ions in the path. In this mode, combining and splitting moves can be done to ions that are sitting next to each other in a storage zone (zone 1, 2, or 3).

\subsection{Error models for our trapped ion VQD}

We take it that following state preparation, with a small probability a given ion is outside the level that defines state $\ket 0$; we consider this the primary noise in the qubit initialisation.  The error is modelled with parameterised amplitude damping
(\Cref{eq:damp}) with initialisation fidelity as the parameter. In making this simple choice we are effectively assuming that the probability of initialising to a state outside of the qubit basis is negligible. 

As noted earlier, we model the remote entanglement operation as a direct
initialisation to the Bell pair
$\ket{\Psi^+}\equiv(\ket{01}+\ket{10})/\sqrt{2}$ for the relevant ions. We model this by initialising the corresponding qubits to state $\ket{\Psi^+}$, followed by noise.  Following the indications in Ref.~\cite{PhysRevLett.124.110501}, we
take the noise to be primarily phase-type  (\Cref{eq:deph2}) but with a small
component of depolarising noise (\Cref{eq:depol2}). The latter accounts for the non-zero probability associated with even parity states. 

The qubit readout error is modelled as symmetric bit-flip noise, with the same
errors for both the ``0'' and ``1'' outputs. This error model is
straightforwardly described by the bit-flip error in \Cref{eq:bf}. To avoid
scattering during the optical excitation and classification of outcomes, qubit readout can only be performed on a single ion in a zone.

The noise in the logical single-qubit rotations ($Rx(\theta)$ and $Ry(\theta)$)
is modelled as depolarising noise, which captures any unknown interactions
during the gate operations.  The error severity increases as $\theta$ increases, which also corresponds to a longer gate duration.

The dominant source of error in the two-qubit $\cz$ gate is taken to be phase flip noise, which is represented by \Cref{eq:deph2}, with a minor contribution from depolarising noise, described by \Cref{eq:depol2}. Detailed audits of the noise in laser-driven gates are well-documented in the literature, see \eg Ref.~\cite{PhysRevLett.124.110501}; the simple choice of noise we make here could be replaced with a form tailored to any given specific system.

In the model, shuttling movements (including physical $\psw$) do not result in
any specific noise. The noise that occurs during a move is only due to passive
noise, which includes an exponential decay of $T_1$-relaxation modelled by
depolarising noise (\Cref{eq:depolt1}) and a Gaussian decay of
$T_2^*$-relaxation modelled by dephasing noise (\Cref{eq:depht2s}). The main cost of this operation comes from accelerating and decelerating the ions; this effect is captured in the duration of shuttling. 
For example, the shuttling duration increases when some ions in zone 1 are moved to zone 3, being $\sqrt2$ times longer than when they are moved to zone 2.

\subsection{Entanglement distillation on trapped ions}

We now describe the use of this VQD to examine an interesting topic: the relative merits of different various entanglement distillations protocols on two remote ion traps.  
With the help of the virtual device, we can easily identify the best strategy to perform entanglement distillation, and understand how this choice depends on the error assumptions and constraints in the model.

\begin{figure}[hpbt]
    \subfloat[``\textbf{bf}'']{
        \includegraphics[scale=0.65]{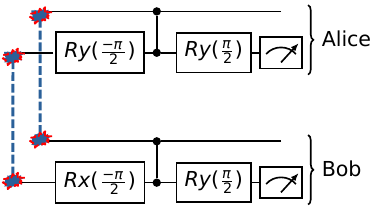}
    }
    \subfloat[``\textbf{ph}'']{
        \includegraphics[scale=0.6]{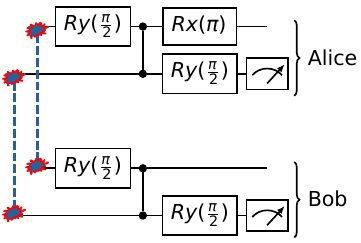}
    }
    \caption{\label{fig:distillation_basic} The canonical entanglement distillation circuits -- between Alice and Bob -- 
        expressed in the native gates of our trapped ion VQD.
        Circuit (a) mitigates bit-flip error and (b) mitigates phase-flip error.  
        We refer to circuit (a) as ``\textbf{bf}'' and (b) as ``\textbf{ph}''.
    The blue dashed lines indicate Bell states shared between Alice and Bob. 
        In each circuit presents two pairs of Bell states that have gone through the same distillation steps;
        the pairs are identical in the ideal case. 
}
\end{figure}
\begin{figure}[hpbt]
    \centering
    \includegraphics[scale=0.9]{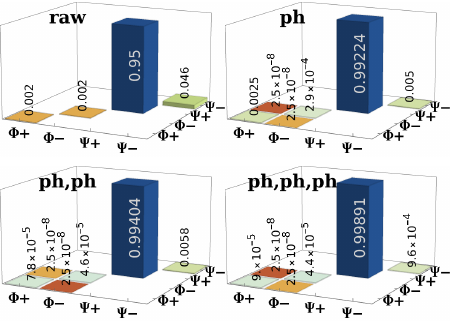} 
    \caption{
        \label{fig:dist_best}
        Density matrices tracking the state through an entanglement distillation involving three rounds of
        phase-flip purification (the best strategy found).
    The state is shown in Bell basis, where 
    $\ket*{\Psi^\pm}=(\ket{01}\pm\ket{10})/\sqrt2$
    and 
    $\ket*{\Phi^\pm}=(\ket{00}\pm\ket{11})/\sqrt2$.
    The primary error source in the raw Bell pair is phase flip, with a probability
    of 0.046 for state $\ket*{\Psi^-}$. Hence, the first round of error correction should be
    phase-flip distillation. After the first round, the phase-flip error is
    reduced, but it remains the dominant source of error. To further reduce the
    error, a second and then ultimately a third phase-flip distillation round is applied. Notably, variants where one round is a bit-flip distillation have inferior final fidelity.
    }
\end{figure}

\begin{figure}[hpbt]
    \centering
    \includegraphics[width=0.9\columnwidth]{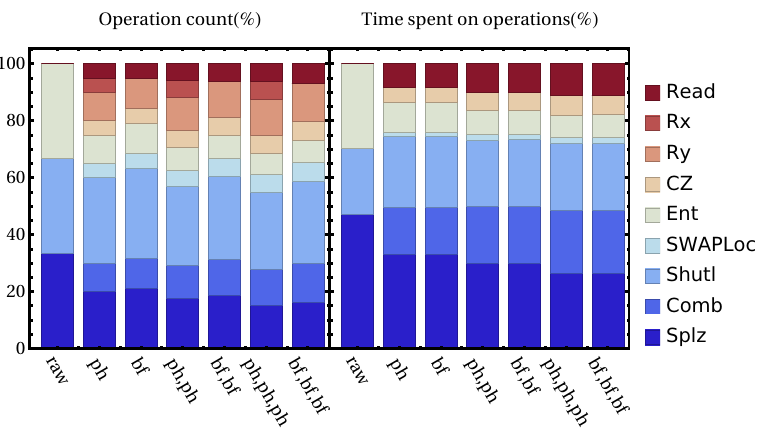} 
\caption{\label{fig:tion_time}
The time profile of the operations involved in the distillation process for a
single node, the Alice node. The left figure represents the percentage of the
total duration time, while the right figure represents the percentage of the
operator count. The overall time required for the process depends on the
strategy employed. Acquiring a raw Bell pair takes 0.1$ms$, a single round of
distillation takes 0.6-0.7$ms$, two rounds of distillation take 1.4-1.7$ms$, and
three rounds of distillation take 3.1-3.8$ms$. There are 112 steps involved for three rounds of phase-flip distillation that is listed in \Cref{app:distillation}. \texttt{Read} indicates readout.
Quantum gates $Rx(\theta),Ry(\theta)$ and $\cz$ are denoted by \texttt{Rx}, \texttt{Ry}, and \texttt{CZ}, consecutively. The moves such as swap, shuttle, combine, and split are indicated by \texttt{SWAPLoc}, \texttt{Shutl}, \texttt{Comb}, and \texttt{Splz}, respectively.
}
\end{figure}
\begin{figure}[hpbt]
    \centering
    \includegraphics[width=0.85\columnwidth]{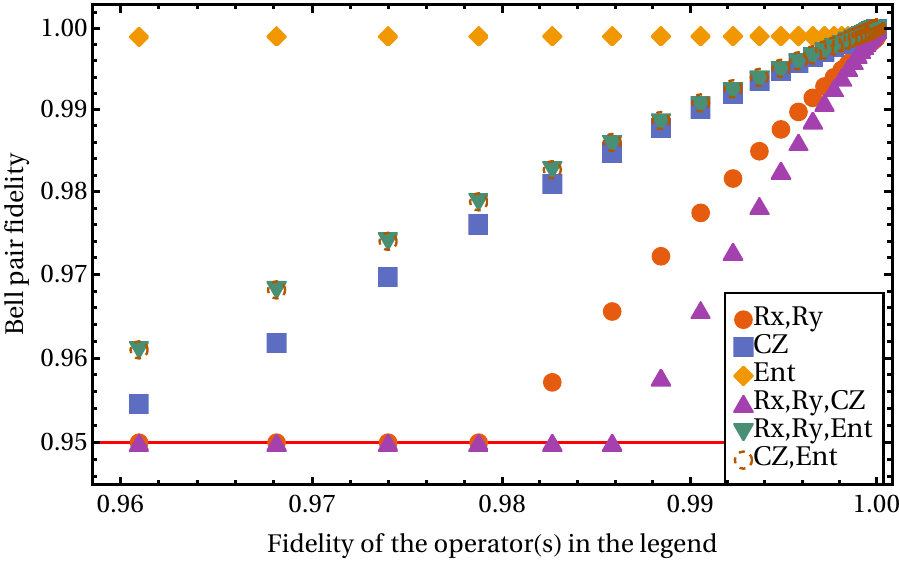} 
\caption{\label{fig:tion_vars}
    Achievable fidelity for a wide variety of VQD variants and protocols. Each point shows the highest Bell fidelity that can be reached by \emph{any} choice of protocol from: `do nothing', to any combination of `bf' and `ph' over one, two or three rounds (see Fig. \ref{fig:distillation_basic} for the circuits). Each family of points corresponds to modifying the default fidelity of certain VQD operations -- as noted in the key -- to the value indicated by the $x$-axis.
    The default relevant fidelities are the following: $\text{Fid}_{Rx,Ry}=0.99999$, $\text{Fid}_{\cz}=0.999$, and $\text{Fid}_{Ent}=0.95$. 
   The red line indicates the default fidelity of the raw Bell pair; when points lie on this line, the gate fidelities are so poor that distillation fails to make any improvement.
}
\end{figure}

In our simulation, we were able to improve the fidelity of a Bell pair from
0.95 to 0.9989. The parameters used in our model are detailed in
\Cref{conf:trapped_ions}. Given such a configuration, the optimal approach
would be to perform three rounds of phase-flip correction, as illustrated in
\Cref{fig:dist_best}.

In order to determine the strategy with the highest yield, we simulate up to
three rounds of entanglement distillation using our VQD.  The
simulation is based on the scenario where there are two traps, each containing
four ions, as illustrated in \Cref{fig:tions}.  In every distillation round, we
run the canonical distillation circuits defined in
\Cref{fig:distillation_basic}.

It is interesting that the best distillation strategy is one that employs exclusively
 phase-flip correction (see \Cref{fig:dist_best}); this of course stems from
the underlying error model. 
As described earlier, motivated by experimental literature our VQD assigns by far the most severe noise to Bell state generation, and following that, to two-qubit $\cz$ operations. We can confirm from
\Cref{fig:tion_time} that these components are the primary contributors to the process, and thus the primary sources of noise. Given that dephasing noise is the
predominant error present in both operations, it is intuitive that phase-flip correction techniques
are the most appropriate.

To gain a deeper understanding of each operator's role in the distillation
process, we modified the fidelity of some operators and selected the Bell pair
with the highest fidelity, creating a large number of VQD variants as shown in \Cref{fig:tion_vars}. In each case, we report the best attainable Bell state fidelity. The figure highlights the significance of
 single-qubit operations: distillation fails entirely when the single-qubit rotations have
fidelities below 0.98. Remarkably, one can make some improvement to the Bell pair even with
noisy $\cz$ gates; the threshold is around 0.96.  
Overall, our VQD simulations demonstrate the robustness of standard entanglement distillation circuits, which is an encouraging observation for the prospects of trapped ions distributed QC.

\section{NV-centre qubits}

We now introduce our second VQD: an NV-centre with its characteristic star topology of two-qubit interactions. We will use this VQD to explore a relatively recent theoretical proposal for exploring a BCS model with a quantum computer.

\subsection{NV-centre physical system}

The nitrogen-vacancy centre (NV-centre) is a common point defect in diamonds, consisting substitutional nitrogen atom next to a vacant lattice site. The
point defect can have a neutral charge (NV${}^0$) or negative charge (\nv) with
triplet spin ($S=1$). In this quantum system, two kinds of qubits are commonly
used for quantum computing purposes: the \nv electron spin and proximal
\cs nuclear spins (abundance of around 1\%).  Both the electron and nuclear
spins are spin-1 systems, two energy levels are deemed to make up a qubit: for example, $m_s=0$
encodes $\ket0$, and $m_s=-1$ encodes $\ket1$. The nitrogen (\ns) nucleus itself is spin-bearing, and is therefore another potential qubit. If we choose to model this in a given VQD, it is trivial to do so by modifying the gate characteristics of one of the `satelite' qubits accordingly.  NV-centre devices of this general kind have been widely studied; for example, researchers at the University of Delft have a considerable body of work including Refs.~\cite{bernien,cramer,abobeih}.

The NV-centre is a central-spin system with the electron spin as the centre and
it couples the nuclear spins via hyperfine interactions, while the nuclear spins themselves are coupled to each other via weak dipolar interactions~\cite{abobeih2019atomic}.  The nuclear-electron coupling strengths are varied depending on the nuclear spin relative position to the electron spin. Such coupling shifts are used to detect, isolate, and coherently control the individual \cs qubit via the central spin. In particular, the hyperfine splittings allow conditional control of the nuclear spins depending on the electron spin state, allowing the realisation of two-qubit gates. Two-qubit gates between nuclear spins are typically realised by a sequence of operations between the nuclear and the electron spins~\cite{taminiau2014universal,abobeih2022fault}. Single-qubit rotations on the electron spin are realised by microwave pulses, while the nuclear-spin rotations are realised by RF-pulses~\cite{abobeih2022fault}. However, in such a system, the spins are continuously coupled with each other; therefore,  elective control operations typically consider various echos sequences that isolate unwanted interactions.

\subsection{NV-centre architecture and native operations}

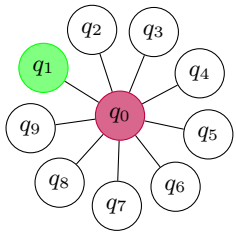
\begin{figure}[thp]
    \centering
\begin{tikzpicture}
    \def \radius {1.4cm}
\node[draw=purple, circle,fill=purple!60] at (360:0mm) (ustar) {$q_0$};
\node[draw=green, circle, fill=green!50] at ({108+40}:\radius) (un) {$q_1$};
\foreach \i [count=\ni from 0] in {2,3,4,5,6,7,8,9}{
    \node[draw=black, circle] at ({108-\ni*40}:\radius) (u\ni) {$q_{\i}$};
  \draw (ustar)--(u\ni);
}
  \draw (ustar)--(un);
\end{tikzpicture}
    \caption{
        A central spin system (CSS) topology describes the architecture of an NV-centre qubit device. 
        In this example, there are 10 qubits with indices $\{q_0,\dots,q_9\}$. The purple, green, and white nodes
correspond to \nv, \ns, and \cs, respectively. The electron spin always sits in the center
with qubit order 0 ($q_0$). The nuclear \ns spin is numbered as 1 ($q_1$) --- if modelled.
Conditional operations can only be done conditioned on the state of qubit $q_0$.
}
\label{fig:star}
\end{figure}

The NV-centre quantum system has star-shaped connectivity with the electron spin lying in the centre, as shown in \Cref{fig:star}. Therefore, direct two-qubit
operations can only be done between the electron and a nuclear spin.   

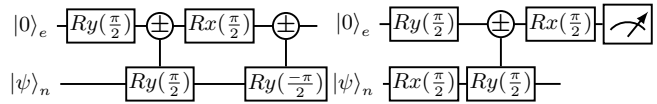
\begin{figure}[hbtp]
\hspace{-5mm}
\resizebox{0.53\columnwidth}{!}{
\begin{tikzpicture}[thick]
\tikzset{
operator/.style = {draw,fill=white,minimum size=1.5em,inner sep=1pt},
phase/.style = {draw,fill=white,shape=circle,minimum size=5pt,inner sep=0pt}
}
\matrix[row sep=3mm, column sep=1mm] (circuit) { 
\node (q1) {${\ket{0}}_e$};
&[-1mm]
&\node[operator] (R11) {$Ry(\frac{\pi}{2})$};
&[-2.5mm]\node[phase] (C12) {$\bm\pm$};
&[-2.5mm]\node[operator] (R13) {$Rx(\frac{\pi}{2})$};
&[-2.5mm]\node[phase] (C14) {$\bm\pm$};
&[-2.5mm]\coordinate (end1);\\
\node (q2) {$\ket{\psi}_n$};
&
&
&\node[operator] (R22) {$Ry(\frac{\pi}{2})$};
&
&\node[operator] (R24) {$Ry(\frac{-\pi}{2})$};
&\coordinate (end2);\\
};
\begin{pgfonlayer}{background}
\draw[thick] (q1) -- (end1)
(q2) -- (end2)
(C12) -- (R22)
(C14) -- (R24);
\end{pgfonlayer}
\end{tikzpicture}
}%
\hspace{-4mm}
\resizebox{0.55\columnwidth}{!}{
\begin{tikzpicture}[thick]
\tikzset{
operator/.style = {draw,fill=white,minimum size=1.5em,inner sep=1pt},
phase/.style = {draw,fill=white,shape=circle,minimum size=5pt,inner sep=0pt},
meter/.append style={draw, inner sep=5, rectangle, font=\vphantom{A}, minimum width=20, line width=.8,
 path picture={\draw[black] ([shift={(.1,.2)}]path picture bounding box.south west) to[bend left=50] ([shift={(-.1,.2)}]path picture bounding box.south east);\draw[black,-latex] ([shift={(0,.1)}]path picture bounding box.south) -- ([shift={(.3,-.1)}]path picture bounding box.north);}}
}
\matrix[row sep=3mm, column sep=1mm] (circuit) { 
\node (q1) {${\ket{0}}_e$};
&[-1mm]
&\node[operator] (R11) {$Ry(\frac{\pi}{2})$};
&\node[phase] (C12) {$\bm\pm$};
&[-2.5mm]\node[operator] (R13) {$Rx(\frac{\pi}{2})$};
&\node[meter] (M14){};\\
\node (q2) {$\ket{\psi}_n$};
&
&\node[operator] (R21) {$Rx(\frac{\pi}{2})$};
&\node[operator] (R22) {$Ry(\frac{\pi}{2})$};
&\coordinate (end2);\\
};
\begin{pgfonlayer}{background}
\draw[thick] (q1) -- (M14)
(q2) -- (end2)
(C12) -- (R22);
\end{pgfonlayer}
\end{tikzpicture}
}
    \caption{\label{fig:nvinitmeas} Initialisation via swap (left) and measurement (right) on a nuclear 
    \cs spin. Indices $q_0$ and $q_j$ indicate \nv electron spin and \cs nuclear spin, respectively.
    The electron spin needs to be initialised to state $\ket0$, while $\ket\psi$ indicates an 
    arbitrary state. {The controlled($\pm$)-$Ry$ gate, notated as $\mathit{CRy^{\pm}}$, is a native NV-center two-qubit operator that is defined in \Cref{eq:conditional_rots}}.  
}
\end{figure}

Direct initialisation and (destructive) readout of qubits are restricted to the \nv
electron spin.  Both processes are done via resonant optical excitation with
certain frequencies that are obtained experimentally. Therefore, nuclear spin
initialisation and readout -- as well as nondemolition measurement -- are done via the \nv electron spin~\cite{abobeih}. For instance,
the circuits in \Cref{fig:nvinitmeas} show initialisation
and measurement of a nuclear spin.

The native operations of our NV-center virtual device comprise single Pauli rotations on the electron and nuclear spins and conditional Pauli rotations between the electron and any given nuclear spin. The single rotations comprise
$\{ Rx_q(\theta),Ry_q(\theta),Rz_q(\theta)\}$,
where $q$ indicates an electron or a nuclear spin. 
The $Rz(\theta)$ rotations are implemented virtually in practice~\cite{abobeih2022fault}, which is done by tracking and accumulating the phase that will be incorporated into the phase of the consecutive pulse~\cite{mckay2017efficient}.
The two-qubit gates comprise conditional rotations
\begin{equation}
    \{\mathit{CRx}^\pm_q(\theta),\mathit{CRy}^\pm_q(\theta)\},
\end{equation}
conditioned to the \nv electron state,  
where $q$ denotes a nuclear \cs spin. The \nv state determines 
the rotation direction, namely  
\begin{equation}\label{eq:conditional_rots}
\begin{aligned}
    \mathit{CRx}^\pm_q(\theta)&\coloneqq\ketbra0_{q_0} \otimes Rx_q(\theta)+\ketbra 1_{q_0} \otimes Rx_q(-\theta)\\
    \mathit{CRy}^\pm_q(\theta)&\coloneqq\ketbra0_{q_0} \otimes Ry_q(\theta)+\ketbra 1_{q_0} \otimes Ry_q(-\theta)
\end{aligned}
\end{equation}
where $q$ is an index of a \cs spin.

\subsection{Error models for our NV-centre VQD}\label{sec:em_nvc}

A high-fidelity electron spin initialisation comprises two stages: optical pumping to obtain \nv from NV${}^0$, and initialisation of \nv to spin $m_s=0$ (state $\ket0$) that is indicated with photon emissions~\cite{robledo2011high}.
We model this by amplitude damping (\Cref{eq:damp}), in which the parameter directly denotes initialisation infidelity.

The \nv electron qubit readout is modelled with a projective measurement in the
computational basis $M_0=\{\ketbra0,\ketbra1\}$.  Recall that, qubit readout is
performed by applying the readout frequency that is resonant with $m_s=0$
excitation: detecting a photon means state $\ket 0$ (or a dark count error) while  seeing no photon means state
$\ket 1$ (or a loss error).  
Our model considers only errors caused by photon loss, as the probability of dark count is typically very low. Hence, the
readout error is asymmetric: given an input state $\ket 1$ measurement is correctly ``1'', but input $\ket 0$ can lead to either ``0'' or ``1''. 

For this, we model the measurement error with an inverted amplitude damping (\Cref{eq:damp}), which
``decays'' the qubit toward state $\ket 1$,
\begin{equation}
    \begin{quantikz}[column sep=0.2cm]
        \lstick{$q_0$} & \gate{X} & \gate{\mathit{damp(p)}} & \gate{X} & \meter{\texttt{0/1}} & \gate{\mathit{damp}(1)}&\qw,
\end{quantikz}
\end{equation}
where $p$ is measurement bias; this gives an appropriate rate of incorrect ``1'' results on input $\ket{0}$. The final damping fully resets the qubit 
to $\ket 0$; this represents the ultimate effect of the repeated measurement process which effectively drives the state to $\ket 0$. Therefore, 
the $p$ used here should be the net (rather than per-measurement) probability that an actual $\ket{0}$, pre-measurement, is recorded as a `1'.

It is important to mention that our measurement model oversimplifies some of the processes occurring in the experiment. Specifically, incorrect assignment of the measurement outcome has the potential to completely dephase the nuclear spins, given that their evolution frequencies depend on the NV electron state. Additionally, the dark count, while potentially very low (<0.1\%), is still an important consideration in the context of repeated measurements, as is typical in actual practice. Despite these complexities, we proceed under the assumption that the fidelity of our measurement model adequately reflects the statistical distribution of such outcomes. Consequently, the assumed measurement fidelity in this model is based on the premise of repeated measurements having been performed.

The physical quantum gates experience standard dephasing and depolarising forms
of noise, while the virtual gate $Rz(\theta)$ can only experience dephasing noise. In practice, the virtual gates are often modelled to be perfect.
Noisy single gate rotations $Rx(\theta)$ and $Ry(\theta)$ are followed by
single-qubit depolarising (\Cref{eq:depol1}) and dephasing (\Cref{eq:deph1})
noise. Noisy two-qubit gates $\mathit{CRx}^\pm$ and $\mathit{CRy}^\pm$
are followed by two-qubit depolarising (\Cref{eq:depol2}) and dephasing
($\Cref{eq:deph2}$) noise. The proportion of depolarising and dephasing
noise are adjustable as per the method discussed in \Cref{sec:standard_error}.

Noise acting on the passive qubits involves free induction decay corresponding to $T_1$
(\Cref{eq:depolt1}) and $T_2$ (\Cref{eq:depht2}), and cross-talk in the form of
    $ZZ$-coupling between all possible pairs of nuclear \cs qubits. Our VQD by default incorporates the assumption that dynamical decoupling is constantly implemented on the passive qubits; 
    this is supported by the common experiment practice in order to achieve long-coherence qubits~\cite{abobeih2018one,bradley2019ten} and to perform decoherence-protected gates~\cite{xu2012coherence,abobeih2022fault}.
    Therefore, we use the $T_2$ as the phase decay rate, and gate implementation is done
    serially. The $ZZ$-coupling captures the slow (a few Hertz) but
    constant unwanted entangling dipolar interaction among the nuclear qubits.

\subsection{BCS model simulation on NV-centre}

As noted in the previous sections, the NV centre has a characteristic topology wherein multiple nuclear spin qubits can interact with a common core, but not with each other directly. A recent paper, ``Digital quantum simulation of the BCS model with a central-spin-like quantum processor''~\cite{ruh2022digital} explores the utility of exactly such a topology for the dynamics Bardeen–Cooper–Schrieffer (BCS) models -- the authors note that NV centres are a natural match to their required geometry. This affords an intriguing opportunity to test our NV Centre VQD -- through the use of the VQD we can determine how experimentally challenging it would be to perform an accurate simulation on a modest scale.

To maximise the prospects for the approach, we opt to synthesise compact circuits that are
propagators of the BCS Hamiltonian; for this we use a recently discussed \textit{subspace compilation} method~\cite{meister2022exploring,gustiani2022exploiting}.
Then, we translate the resulting circuit into the native operations
of our NV-centres. Finally, we examine the simulation performance under
various scales of noise.

We follow the prescription of \citeauthor{ruh2022digital}\cite{ruh2022digital}, who explain that a relevant BCS model can be connected to central spin systems with the Hamiltonian, 
\begin{equation}\label{eq:mainBCS}
    H_\text{BCS}(t)=-g(t)\sum_{q=0}^{n-1}{\epsilon_q H_q(t)}+g(t)L^z+g(t)(L^z)^2,
\end{equation}
where $\epsilon_q$ is energy orbital, $g(t)$ is the magnetic field quench.
In the context of interest for \citeauthor{ruh2022digital}, the Gaudin Hamiltonians $H_q$ can be defined as
\begin{equation}
    H_q(t)=\sum_{j=0,j\neq q}^{n-1}
    \frac{\vec\sigma_q\cdot\vec\sigma_j}{2(\epsilon_q-\epsilon_j)}+\frac{\sigma_q^z}{g(t)},
\end{equation}
where $\vec\sigma\equiv(\sigma^x,\sigma^y,\sigma^z)$  is a Pauli vector.

\begin{figure}[hbtp]
    \includegraphics[width=\columnwidth]{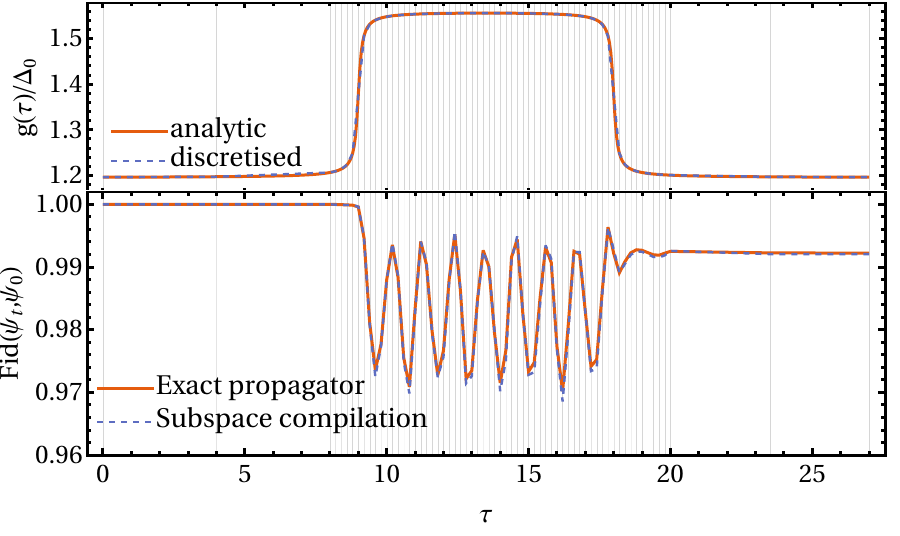}
    \caption{\label{fig:quench}A quench is driven by varying $g$ within $H_\text{BCS}$ as shown in upper panel. The corresponding `return probability' Fid$(\psi_t,\psi_0)$ defined in \Cref{eq:return} is shown in the lower panel. The horizontal axis ($\tau$) is scaled time to the unit energy 
     as per Ref.~\cite{ruh2022digital}. 
        The grey lines indicate time discretisation comprising 64 partitions; thus involving 64 propagators.
        We use the exact initial state of $H_\text{BCS}$ at $t=0$.
        The propagators are computed exactly using \texttt{Mathematica} function
        \texttt{MatrixExp},
        and approximated with circuits obtained via the subspace compilation method introduced in~\cite{meister2022exploring,gustiani2022exploiting}.  
}
\end{figure}

\citeauthor{ruh2022digital} note that the model is integrable when parameters are suitable constants, but that by considering a time-varying $g(t)$ one can realise a quench. They consider the scenario where $g$ switches (smoothly, via a compact S-curve) from a base value to an increased value, before then switching back to base. We follow their implementation, see~\Cref{fig:quench}. The authors monitor the quantity called the return probability which they denote $\mathcal{R}_\text{exact}$ and we write 
as follows
\begin{equation}\label{eq:return}
    \text{Fid}(\psi_t,\psi_0)
    =|\bra{\psi_0}\mathcal T[e^{-\frac{i}{\hbar}\int_{t'=0}^{t'=t}H_\text{BCS}(t')dt'}]\ket{\psi_0}|^2,
\end{equation}
as it is the state's fidelity with respect to the initial state at $t=0$ ($\psi_0$).
Here $\mathcal T$ is the time-ordering operator. The initial state
$\ket{\psi_0}$ is the ground state of the $H_\text{BCS}$ at $t=0$ that is obtained exactly. With this setup, we can observe the oscillation around the 
quench as shown in the second figure of \Cref{fig:quench}. The simulation is done for $n=5$ qubits and the orbital energy levels are defined as simple non-interacting harmonic oscillators as per reference~\cite{ruh2022digital}.

To simulate dynamics of a BCS model, as the authors of~\cite{ruh2022digital} note the state at $t$ is evolved with respect to that at $t_0$ by an operator,
\[
U(t,t_0)=U(t=t_m,t_{m-1})...U(t_1,t_0)
\]
with
\[
U(t_j,t_{j-1})\approx e^{-i/\hbar\,\,H(t_{j-1})\,\,(t_j-t_{j-1})}
\]
provided that $t_j-t_{j-1}$ is sufficiently small. Each $U(t_j,t_{j-1})$ must then be expressed as an NV-compatible circuit. A canonical solution is 
Trotterisation as employed in~\cite{ruh2022digital} to express the Gaudin terms,
the most complex terms in $H_\text{BCS}$. Even though the Gaudin term
employs only one SWAP gate to have star-shaped connectivity, the gate count in
Trotterisation increases severely with the desired accuracy; we find this can require around one
thousand gates for {
each propagator $e^{-i \Delta t H_\mathit{BCS}(t)}$ to obtain fidelity accuracy at least $10^{-3}$}. 
Consequently, if hardware noise were present at any meaningful level, the output would be almost totally corrupt.

\begin{figure}[!b]
\centering
    \subfloat[Propagators gate counts]{
        \includegraphics[width=0.8\columnwidth]{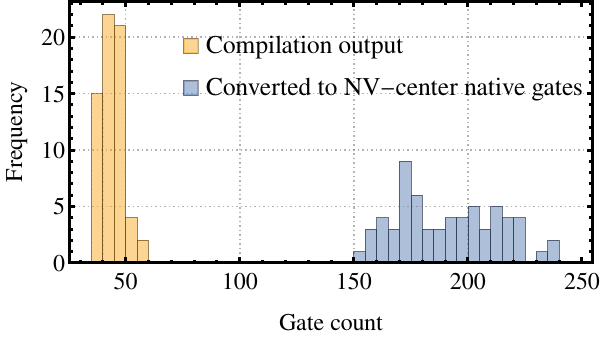}
    }\\ 
\subfloat[Controlled-z rotation expressed in native NV-center gates]{
\hspace{-1em}
\begin{adjustbox}{max width=0.5\textwidth}
\begin{tikzpicture}[thick,baseline={([yshift=-.5ex]current bounding box.center)},vertex/.style={anchor=base}]
\tikzset{
operator/.style = {draw,fill=white,minimum size=1.5em,inner sep=1pt},
phase/.style = {draw,fill=black,shape=circle,minimum size=5pt,inner sep=0pt}
}
\matrix[row sep=3mm, column sep=1mm] (circuit) { 
\node (q1) {$q_0$};
&\node[phase] (C12) {};
&\coordinate (end1);\\
\node (q2) {$q_j$};
&\node[operator] (R22) {$Rz(\theta)$};
&\coordinate (end2);\\
};
\begin{pgfonlayer}{background}
\draw[thick] (q1) -- (end1)
(q2) -- (end2)
(C12) -- (R22); 
\end{pgfonlayer}
\end{tikzpicture}
\hspace{-3mm}
=
\hspace{-3mm}
\begin{tikzpicture}[thick,baseline={([yshift=-.5ex]current bounding box.center)},vertex/.style={anchor=base}]
\tikzset{
    operator/.style = {draw,fill=white,minimum size=1.5em,inner sep=1pt},
phase/.style = {draw,fill=white,shape=circle,minimum size=5pt,inner sep=0pt}
}
\matrix[row sep=0.2cm, column sep=1mm] (circuit) { 
\node (q1) {$q_0$};
&\node[operator] (R11) {$Ry(\frac{-\pi}{2})$};
&[-3mm]\node[phase] (C12) {$\bm\pm$};
&[-3mm]\node[operator] (R13) {$Ry(\frac{\pi}{2})$};
&[-4mm]\node[phase] (C14) {$\bm\pm$};
&\node[operator] (R15) {$Rx(\frac{\theta}{2})$};
&\node[phase] (C16) {$\bm\pm$};
&\coordinate (end1);\\
\node (q2) {$q_j$};
&
&\node[operator] (R22) {$Rx(\frac{\theta}{2})$};
&
&\node[operator] (R24) {$Ry(\frac{-\pi}{2})$};
&\node[operator] (R25) {$Rx(\frac{-\theta}{2})$};
&\node[operator] (R26) {$Ry(\frac{\pi}{2})$};
&\coordinate (end2);\\
};
\begin{pgfonlayer}{background}
\draw[thick] (q1) -- (end1)
(q2) -- (end2)
(C12) -- (R22) 
(C14) -- (R24) 
(C16) -- (R26) ;
\end{pgfonlayer}
\end{tikzpicture}
\end{adjustbox}
}

\subfloat[Parameterised swap expressed in native NV-center gates]{
\begin{adjustbox}{max width=0.5\textwidth}
\hspace{-1em}
\begin{tikzpicture}[thick,baseline={([yshift=-.5ex]current bounding box.center)},vertex/.style={anchor=base}]
\tikzset{
crossx/.style={path picture={
\draw[thick,black,inner sep=0pt]
(path picture bounding box.south east) -- (path picture bounding box.north west) (path picture bounding box.south west) -- (path picture bounding box.north east);
}}
}
\matrix[row sep=0.2cm, column sep=1mm] (circuit) { 
\node (q1) {$q_0$};
&\node[crossx] (C12) {};
&\coordinate (end1);\\
\node (q2) {$q_j$};
&\node[crossx] (C22) {};
&\coordinate (end2);\\
};
\begin{pgfonlayer}{background}
    \draw[thick,shorten >=-4pt,shorten <=-4pt](C12)--(C22) node [midway,right]{$\theta$};
\draw[thick] (q1) -- (end1)
(q2) -- (end2);
\end{pgfonlayer}
\end{tikzpicture}
=
\begin{tikzpicture}[thick,baseline={([yshift=-.5ex]current bounding box.center)},vertex/.style={anchor=base}]
\tikzset{
    operator/.style = {draw,fill=white,minimum size=1.5em,inner sep=1pt},
phase/.style = {draw,fill=white,shape=circle,minimum size=5pt,inner sep=0pt}
}
\matrix[row sep=0.2cm, column sep=1mm] (circuit) { 
\node (q1) {$q_0$};
&
&
&\node[phase] (C13) {$\bm\pm$};
&
&\coordinate (end1);\\
\node (q2) {$q_j$};
&\node[operator] (R21) {$Ry(\frac{\pi}{2})$};
&\node[operator] (R22) {$Rx(\frac{-\theta}{2})$};
&\node[operator] (R23) {$Rx(\frac{\theta}{2})$};
&\node[operator] (R24) {$Ry(\frac{-\pi}{2})$};
&\coordinate (end2);\\
};
\begin{pgfonlayer}{background}
\draw[thick] (q1) -- (end1)
(q2) -- (end2)
(C13) -- (R23);
\end{pgfonlayer}
\end{tikzpicture}   
\end{adjustbox}
}
    \caption{\label{fig:bcsgates}
        The resulting gate count of subspace compilation of 
        $H_\text{BCS}$ propagators before and after conversion to the 
        native NV-centre gates are shown in (a). The gate pool has a star
        topology with electron spin $q_0$ as the central spin: 
        $\{Rz_{q_k}(\theta),C_{q_0}Rz_{q_j}(\theta),\psw_{q_0,q_j}^\theta\}$,
        where 
        $k\in\{0,\dots,n-1\}$ and $j\in\{1,\dots,n-1\}$.
        The controlled-z rotations $C_{q_0}Rz_{q_j}(\theta)$ are converted
        to the native NV-centre gates by circuit (b). The parameterised
        swaps are expressed in the native NV-centre gates by circuit (c).
        The controlled($\pm$)-rotations are defined in \Cref{eq:conditional_rots}. 
    }
\end{figure}
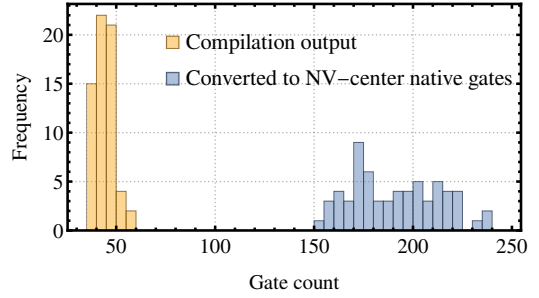
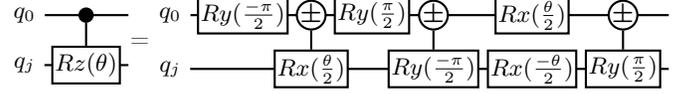
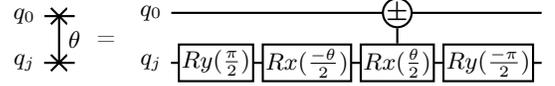

As an alternative, we can use \textit{ab initio circuit synthesis} methods to find circuit realisations. Moreover, the present case is especially suitable as it allows us to take advantage of the
block-diagonal form of $H_\text{BCS}$ --- we can employ a recent subspace compilation method introduced
in~\cite{meister2022exploring,gustiani2022exploiting}. Here, an adaptive variational procedure synthesises a circuit to closely mimic a defined target; the `subspace' qualifier means that the target need only be matched \textit{within} a defined subspace of interest and is free to diverge arbitrarily elsewhere.  We can safely permit this freedom since, by construction, $H_\text{BCS}$ is spin-preserving and thus block diagonal,
because it only rearranges the energy levels that are occupied by the Cooper
pairs~\cite{ruh2022digital}.  We compile each $H_\text{BCS}$ propagator in the
subspace that preserves the spin of the
initial state (that of Hamming weight one).

This initial compilation is not to the NV-native gates, but rather to a gate set that naturally corresponds to the Hamiltonian structure. As one might hope, this produces very compact propagators with gate counts 35-70, as shown in
\Cref{fig:bcsgates}.
Specifically, we set the gate pool with a star topology:
\[\{Rz_{q_k}(\theta),C_{q_0}Rz_{q_j}(\theta),\psw_{q_0,q_j}^\theta\},\]
where
$k\in\{0,\dots,n-1\}$ and $j\in\{1,\dots,n-1\}$ and $q_0$ is set to be the
electron \nv spin. Gate $C_{q0}Rz_{q_j}(\theta)$ is controlled-$z$ rotation
with the controlled qubit is the electron spin $q_0$. Gate
$\psw^\theta=\exp{(-i\frac{\theta}{2}(XX+YY+ZZ))}$ is parameterised swap gate,
where the full swap is achieved when $\theta=m\pi$ for any integer $m$.  Those
gates are chosen for their Hamming weight-preserving property. 
The resulting circuits are then converted into the native NV-centre gates as shown in \Cref{fig:bcsgates}. They then typically include about $200$ gates, still vastly more compact than the direct Trotter approach (and moreover, propagators $U(t_m,t_{m-1})$ can correspond to substantial time periods $t_m-t_{m-1}$ when the function $g(\tau)$ is not changing appreciably). 

\begin{figure}[htp]
    \includegraphics[width=\columnwidth]{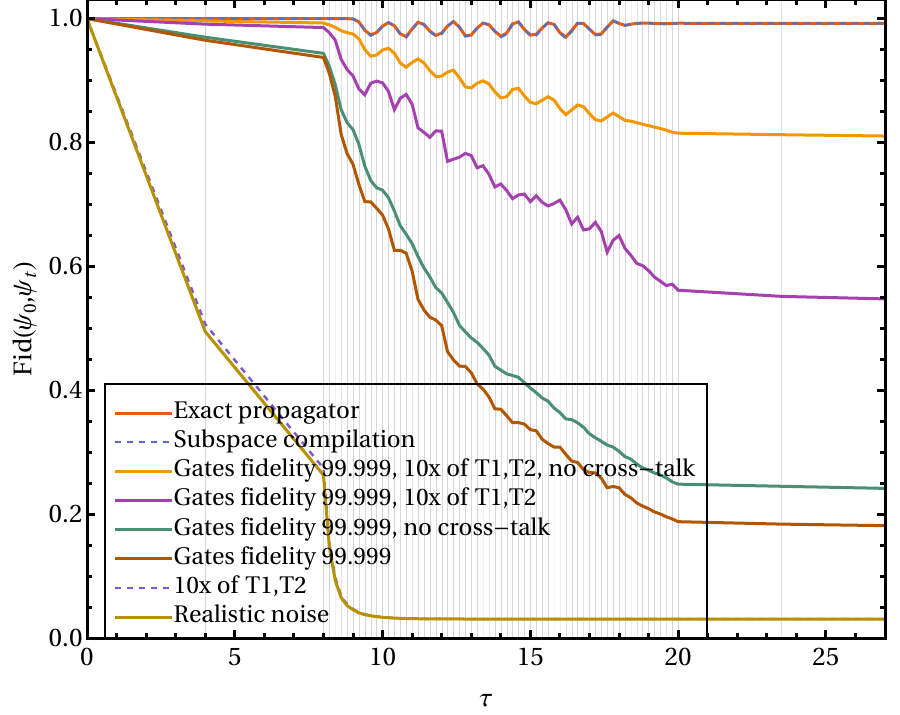}
    \caption{\label{fig:bcsall}
        Simulation of $H_\text{BCS}$ dynamics for various serveries of NV-centre noise.
        There are 64 propagators with the scaled time unit ($\tau$) as per Ref.~\cite{ruh2022digital}.
        The initial state is set to be perfect and exact at $t=0$. The realistic
        noise configuration is given in \Cref{conf:nvc} in the Appendix. Various scale of noise 
        is tried out as shown in the figure; the value on the
        realistic noise setting is used when not mentioned. See \Cref{fig:quench}
        as a reference to the quench function $g(\tau)$. 
    }
\end{figure}

Having thus expressed the complete sequence of propagators using the native NV-centre gates, we now can
estimate the dynamics of $H_\text{BCS}$ on our NV-centre VQD. First, we
perfectly prepare the initial state as the exact $H_\text{BCS}$ ground state for
$t=0$.  We apply the noisy form of the propagators and
calculate the return probability (\Cref{eq:return}) on each time step. Note that, as we assume a continual dynamical decoupling sequence is
applied to the passive qubits, our circuits are implemented in serial manner; this has the implication that they suffer more noise. We measure the performance on various scales of gate noise, with the results shown in 
\Cref{fig:bcsall}.

\Cref{fig:bcsall} shows a poor simulation performance on the realistic noise
setting, even for the first propagator. That is expected since the first
propagator comprises 202 gates -- much larger than the state-of-the-art complex
experiment~\cite{abobeih2022fault}.  Interestingly, the oscillation is already
observed by setting the gates' fidelities to 0.99999 (see the
maroon line in Figure). Moreover, a clear quasi-periodic oscillation manifests when more aspects are improved, \ie high gate fidelities, longer coherence time and the absence of crosstalk  (see the orange line).

The sobering conclusion is that even the ultra-weak nuclear-nuclear crosstalk, if left unmitigated, is sufficient to significantly degrade the model. We would anticipate that increasing the number of qubits to increase the complexity of the isomorphic BCS model, would correspondingly amplify these issues. However, we have not explored error mitigation strategies such as zero-noise extrapolation or quasi-probability constructions~\cite{temme2017error,li2017efficient}, let alone the more recent methods~\cite{cai2022quantum}, and these might alleviate the fidelity requirements very significantly -- an interesting topic for future study, and (needless to say) very straightforward to investigate with the VQD tool.

\section{Neutral atoms}
\subsection{Neutral atoms physical system}
Our neutral atom VQD is based on experimental hardware platforms that exploit quantum registers composed of arrays of individually trapped neutral atoms~\cite{Morgado21}. Each atom encodes a single qubit for alkali atoms such as Rb or Cs by using hyperfine ground-state transitions with microwave frequency separations, whilst for alkaline earth systems using Sr and Yb qubits can be encoded optically via narrow clock transitions using the long-lived intermediate triplet states. Qubit initialisation, state preparation, quantum gates and readout are all performed using lasers providing access to both global and single-site operations. Measurements are performed using fluorescence detection.

A major advantage of the neutral atom platform is scalability, with the ability to prepare various geometric configurations in 1D, 2D and 3D with up to 1000 qubits \cite{barredo2016atom,barredo2018synthetic,Saffman22}. This can be combined with moving optical tweezers to permit dynamic reconfiguration of the qubit register during computation~\cite{bluvstein2022quantum}, providing flexible connectivity.

To engineer coupling between atomic qubits, atoms are excited to Rydberg states, offering strong long-range dipole-dipole interactions. These interactions result in a blockade effect at short range~\cite{lukin2001dipole} whereby only a single atom can be excited to the Rydberg state within a radius of 5-10 $\mu m$. This blockade mechanism can be exploited to implement two- or multi-qubit gates such as Toffoli gates with arbitrary numbers of control qubits ~\cite{isenhower2011multibit,khazali2020fast,levine2019parallel}. Recent experimental demonstrations have shown these gates can be implemented with high fidelity in systems with {$60$} qubits~\cite{Evered23} {in parallel}, making neutral atoms a promising candidate for scalable quantum computing.

\subsection{Neutral atoms architecture and native operations}

\begin{figure}[tp]
\includegraphics[scale=0.5]{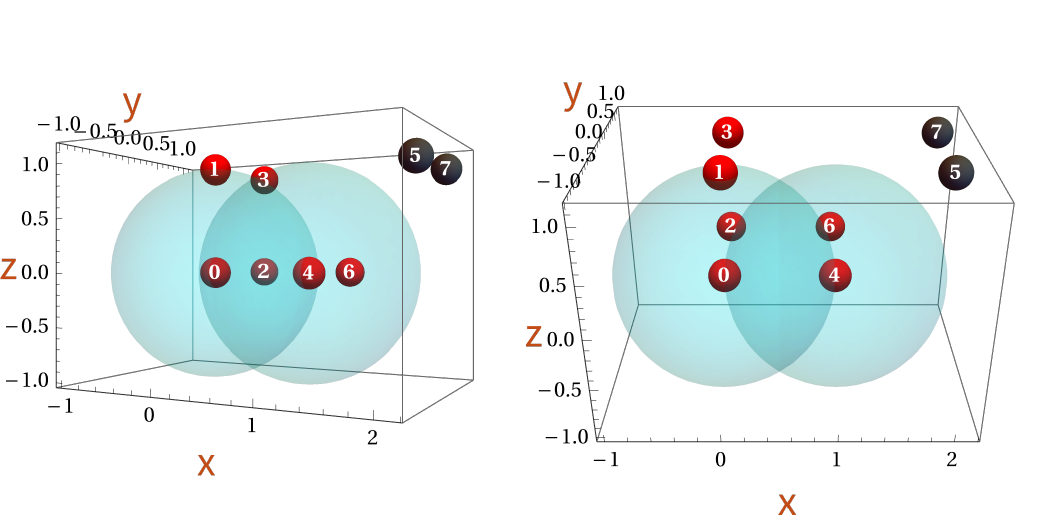}
\caption{
A 3D lattice configuration of a virtual Rydberg neutral atoms.
The red dots indicate the atoms and their corresponding qubit indices. The user can display the blockade radii of selected atoms, as shown for atoms 0 and 4. To perform multi-qubit operations, the atoms must be within the overlapping blockade radii, \eg as in the case of atoms 0 and 4 in the picture, enabling a two-qubit gate between them. The user can also choose to display the last
position of missing atoms, represented by grey atoms. In this example, atoms 5 and 7 are lost to the environment.
}
\label{fig:rydberg3d}
\end{figure}

The architecture of the virtual neutral atoms is entirely configurable by the end-user, who specifies the initial geometry (a list of positions for each atom), blockade radius, and probability of atom loss during reconfiguration. The geometry is arbitrary and can involve a 3D configuration, with visual aids available to assist the user.  For instance, \Cref{fig:rydberg3d} shows a 3D configuration that visualizes two overlapping blockade radii and two atoms that are lost to the environment, which is an additional error mechanism compared to other qubit platforms.

In our virtual neural atoms, it is possible to initialise or reset qubits at any time, either individually or acting on the whole register simultaneously, reflecting the process of trapping and optical pumping of atomic qubits into $\ket 0$.

The qubit readout can be performed concurrently, where it is modelled by performing projective measurement in the computational basis. The measurement process is based on the method used in reference~\cite{nikolov2023randomized}, where in practice, it may also induce atom loss to the environment.

Reconfiguring the register is accomplished by utilising two fundamental move operations: location shift ($\mathit{ShiftLoc_{\mathcal Q}}(\vec v)$) and location swap ($\mathit{SWAPLoc_{p,q}}$). The $\mathit{Shiftloc_{\mathcal Q}}(\vec v)$ operator shifts the position of atoms in set $\mathcal Q$ by a
vector $\vec v$, while the $\mathit{SWAPLoc}_{p,q}$ operator exchanges the positions of atoms $p$ and $q$.

An arbitrary single qubit unitary is achieved by optical control~\cite{xia2015randomized,wang2016single}, by driving two-photon Raman transitions far detuned from an intermediate excited state facilitating rapid population transfer between states $\ket0$ and $\ket1$.  By configuring the phase ($\phi$), two-photon detuning ($\Delta$), and duration of the laser ($t$), an arbitrary single
unitary can be achieved as follows
\begin{multline}
    \label{eq:u}
    U(\phi,\Delta,t)=\\
    \begin{pmatrix}
        \cos(\frac{\tilde\Omega t}{2}) -i\frac{\Delta}{\tilde\Omega}\sin(\frac{\tilde\Omega t}{2})&
        -i\frac{\Omega}{\tilde\Omega}\sin(\frac{\tilde\Omega t}{2})e^{i\phi} & 
        \\[0.5em]
        \mkern-30mu
        -i\frac{\Omega}{\tilde\Omega}\sin(\frac{\tilde\Omega t}{2})e^{-i\phi} &\mkern-20mu 
        \cos(\frac{\tilde\Omega t}{2}) +i\frac{\Delta}{\tilde\Omega}\sin(\frac{\tilde\Omega t}{2})
    \end{pmatrix},
\end{multline}
where $\tilde\Omega=\sqrt{\Omega^2+\Delta^2}$, and $\Omega$ indicates the Rabi frequency. 
For instance, a Hadamard gate can be obtained by operator $U(\phi=0,\Delta=\Omega,t=\pi/\tilde\Omega)$
and a rotation $Rx(\theta)$ can be achieved by operator $U(\phi=0,\Delta=0,t=\theta/\Omega)$.

When two or more Rydberg atoms are excited within a certain distance of each other, they experience a strong interaction that prevents any further excitations within this distance, known as the blockade radius. The interaction leads to the accumulation of a phase that depends on the state of the other atoms within the blockade radius. This phase accumulation is the basis for many multi-qubit gate implementations {
and determines parallel execution of quantum gates}. This process is also known as Rydberg blockade mechanism~\cite{urban2009observation,adams2019rydberg}. 

\begin{figure}[htbp]
\includegraphics[width=4.3cm]{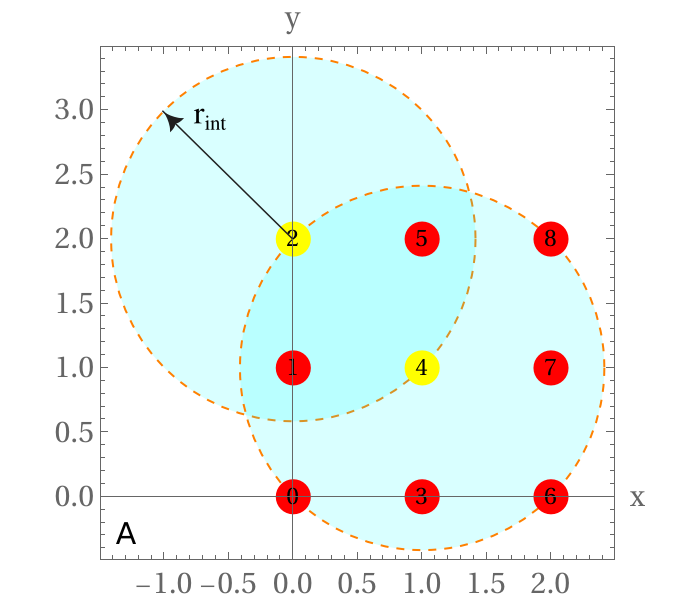}\hspace{-6mm}
\includegraphics[width=4.3cm]{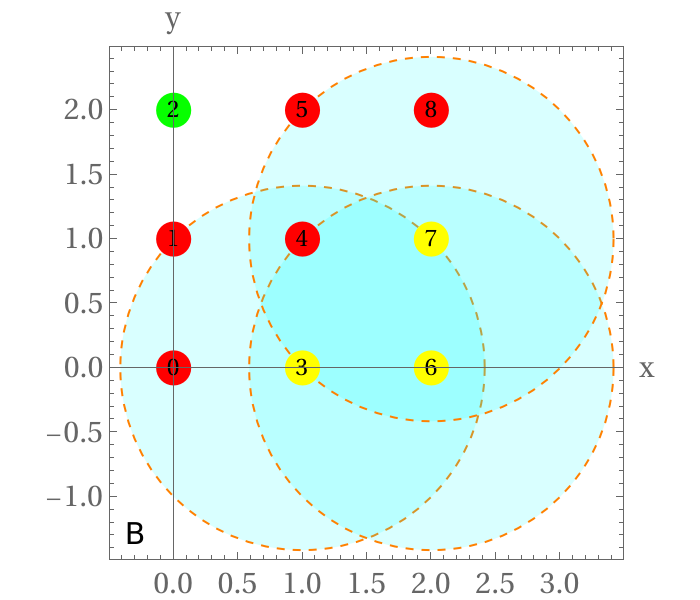}
\caption{\label{fig:blockade}
In panel \texttt{A}, the interaction zones of the yellow atoms (2 and 4), each with radius $r_{int}$, are depicted with the blue area. Since they are within each other's interaction zones, atoms 2 and 4 can execute a two-qubit gate. In panel \texttt{B}, the yellow atoms (3, 4, and 7) are actively interacting, \ie performing a three-qubit gate, thus, the resulting restriction zone encompasses the entire blue area. Consequently, only a gate acting upon atom 2 can be executed in parallel with this three-qubit gate.}
\end{figure}

{
We approximate the Rydberg interaction as a hard 2- or 3-dimensional sphere with radius $R_b$, defining the interaction zone $r_{int}$ (refer to \Cref{fig:blockade}). Consequently, when two or more atoms are within their respective interaction zones, they can interact and execute multi-qubit gates. These interacting atoms collectively form what is termed a \emph{restriction zone}, a union of their interaction zones. This restriction zone inherently limits the parallel execution of other quantum gates within. \Cref{fig:blockade} illustrates these zones, and for more detailed information, see reference~\cite{schmid2023computational}.
Our tool accommodates such a Rydberg blockade mechanism and provides automatic parallelism of quantum operations that adhere to these restriction zones. \Cref{app:parallel_na} provides an explicit example of such parallelism.
}

Based on many proposed and available
techniques~\cite{levine2019parallel,theis2016high,petrosyan2017high,beterov2018fast,beterov2018fast,rasmussen2020single},
we provide the following operations as the native multi-qubit gates in our
 neutral atom VQD.
\begin{align}
    \cz(\phi)&_{p,q}=
    \begin{pmatrix}
        1 & 0 & 0 & 0 \\
        0 & e^{i\phi} & 0 & 0 \\
        0 & 0 & e^{i\phi} & 0 \\
        0 & 0 & 0 & e^{i(2\phi-\pi)} 
    \end{pmatrix}\\
    C_{\mathcal Q}Z_{p}&=
    i(\ketbra*{2^k{-}1}_{\mathcal Q}Z_p +\sum_{j=0}^{2^k{-}2}\ket{j}\bra{j}_{\mathcal Q}\,I_p
    )\\
    C_{p}Z_{\mathcal Q}&=
     i\prod_{j\in\mathcal Q}(\ketbra{0}_p I_j+\ketbra{1}_p Z_j
    ),
\end{align}
where $p$ and $q$ indicate single atom indices, and   $\mathcal Q$ denotes a set of atom indices with cardinality $k$. Parameter $\mathcal\phi$ is a phase provided by the user
that captures the accumulated phase in the process.  To implement the
$\cz(\phi)_{p,q}$ gate, the blockade radii of atoms $p$ and $q$ must overlap in which atoms $p,q$ must lie within. 
The multi-qubit operators $C_pZ_{\mathcal Q}$ acts as the one-control (atom $p$) of the multi-$Z$ operator applied to atoms in $\mathcal Q$,
while the $C_{\mathcal Q}Z_p$ acts as the multi-control (all atoms in $\mathcal Q$) of the $Z$ operator applied to atom $p$.
To simplify the phase-dependence associated with the blockade mechanism, every atom in $\{p\}\cup\mathcal Q$ must be within their overlapping blockade radii, {as illustrated in \Cref{fig:blockade}}.
In our VQD, the blockade condition must be met whenever any of these gates are applied, in order to reflect the blockade mechanism. The question of whether two qubits can interact is therefore a binary one: they can if they are within the radius, otherwise they cannot. Of course, a further extension of this VQD could impose a more nuanced rule whereby phase acquisition rate, and indeed error rates, very continuously with separation. In any case, the end-user's code must take into account any reconfiguration of the register between gate applications.

\subsection{Error models for our neutral atoms VQD}

Device initialisation involves loading a cold ensemble of neutral atoms into the desired register configuration using
optical tweezers~\cite{barredo2016atom,barredo2018synthetic,endres2016atom}. Each atom
encodes a single qubit; the computational basis corresponds to ground-state hyperfine levels and is
initialised through Raman cooling. Imperfect cooling results in some population leakage 
 outside of the computational basis; we can describe this leakage as a
complete-positive and trace-nonincreasing (CPTN) map. Noisy qubit
initialisation is described as a perfect damping (\Cref{eq:damp} with $p=1$) that
resets the qubit to $\ket0$, followed by a CPTN map consisting of operator
\begin{equation}
    L=\begin{pmatrix}
        \sqrt{1-\gamma} & 0 \\
        0 & 1
    \end{pmatrix},
\end{equation}
where $0\leq\gamma\leq 1$ denotes the probability of leakage during initialisation.
Notice that $L^\dagger L\leq I$ and is completely positive, and thus 
it fulfils the requirements of the CPTN map. In practice, the effect is to attenuate the total probability associated with system's state, with the `missing' probability associated with leakage events that are not tracked further within the model.

Readout is achieved by pumping atoms from state $\ket1$ into a stretched state
that enables their distinction from those in state
$\ket0$~\cite{kwon2017parallel}; then a fluorescence image can be taken
to determine the occupancy of the array sites. This method is often accompanied
by atom loss ($\leq 2\%$), which increases with the number of measurements.  We
model this noisy measurement as depolarising noise (\Cref{eq:depol1})
accompanied with a list that tracks accumulated atom loss probability due to
measurement. After a measurement, a pseudo-random number is generated to determine if the atom is still present or lost to the environment.

The error in implementing a single-qubit unitary $U(\phi,\Delta,t)$ in
\Cref{eq:u} mainly comes from the characterisation of noise in Raman
transition, which is dominated by phase-type errors (\Cref{eq:depol1}) with the
error parameter $\alpha=(1-e^{t/2T_2})$ and a small fraction of bit-flip error.
The parameter $\alpha$ corresponds to the probability of a phase flip occurring
during the gate duration $t$. There is a finite bit-flip
error probability which is typically asymmetric, \ie probability of the event $\ket0\mapsto\ket1$
is not equal to the event $\ket1\mapsto\ket0$.

The realization of the two-qubit $\cz(\phi)$ gate utilises the blockade
mechanism in highly excited Rydberg states. The typical error in this mechanism
arises from spontaneous emission from Rydberg states, associated with the possibility of the system decaying to states outside the computational basis. We
model this error using a CPTN map consisting of the following single operator:
\begin{equation}\label{eq:czleak}
    K=\begin{pmatrix}
        1&0&0&0 \\
        0&\sqrt{1-\alpha}&0&0\\
        0&0&\sqrt{1-\alpha}&0\\
        0&0&0&\sqrt{1-\beta}
    \end{pmatrix},
\end{equation}
where $\alpha$ and $\beta$ correspond to the probability of decay to a state
outside the computational basis during the excitation time. Typically, $\alpha$
and $\beta$ have different values, as they correspond to distinct paths in
the process. Notice that $K^\dagger K\leq 1$ and the map is completely positive. 

For multi-qubit gates, we implement a similar noise model as for the two-qubit
gates. Let $\mathcal Q$ be the set of qubits involved in multi-qubit gate
operations, the error during the process is described with a CPTN map 
that comprises the following single operator
\begin{equation}
    M=\bigotimes_{j\in\mathcal Q}M_j\quad
    M_j=\begin{pmatrix}
        1&0\\
        0&\sqrt{1-\alpha}
    \end{pmatrix},
\end{equation}
where $\alpha$ is the leakage probability.

The passive qubits undergo free induction decay,
characterised by $T_1$ and $T_2$ times. 
The $T_1$ time is fundamentally limited by the vacuum lifetime of each atom, denoted as $T_{vac}$. This can be approximated using the relationship $T_1=T_{vac}/N$, where $N$ signifies the number of atoms present in the vacuum chamber. This estimation characterises the \emph{collisional loss} rate, which corresponds to the rate of collision with untrapped background atoms; hence, this rate serves as the upper bound for $T_1$~\cite{saffman2016quantum}. The $T_1$ decay is modelled using depolarising noise (\Cref{eq:depolt1}), while the $T_2$ decay is modelled with dephasing noise (\Cref{eq:depht2}). Both types of noise account for energy loss and information loss resulting from collisions with background atoms.

Reconfiguring the atoms (a `move') is accomplished through the manipulation of optical tweezers. The move operators constitute operations that modify the atoms' locations such as $\mathit{SWAPLoc}$ and $\mathit{ShiftLoc}$.  
The process typically requires a substantial duration in practice, on the order of hundreds of microseconds. Additionally, atoms in motion acquire extra heat -- especially when they are moving fast, resulting in an enhanced dephasing process. The errors acquired during this process constitute the depolarising $T_1$ decay (\Cref{eq:depolt1}) and the enhanced dephasing $\kappa T_2$-decay to account for the thermal effect, namely,
\begin{equation}
deph\left(\frac{1-e^{-\kappa \Delta t}/T_2}{2}\right),
\end{equation}
where the factor $\kappa\geq 1$ intensifies the decay due to the presence of heat. Ideally, $\kappa$ is affected by the speed and the distance of the moves; for simplicity, we set $\kappa$ to be a constant.

\subsection{Graph state construction and Steane code preparation on Neutral Atoms}

\begin{figure}[t]
    \subfloat[1D cluster state.]{
    \includegraphics[width=\columnwidth]{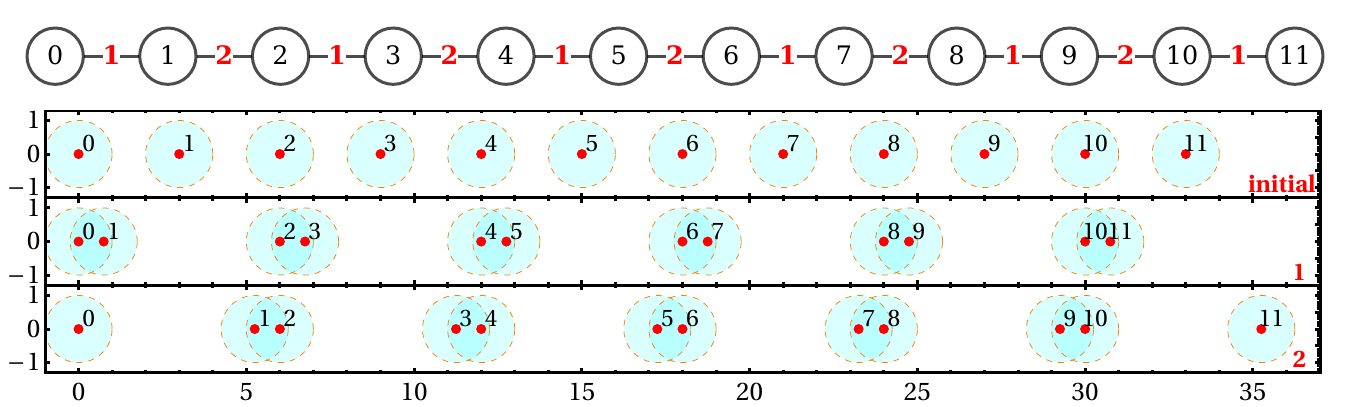}
    \label{fig:graph1d}
    }

    \subfloat[Seven-qubit Steane code logical state.]{
    \includegraphics[scale=0.35]{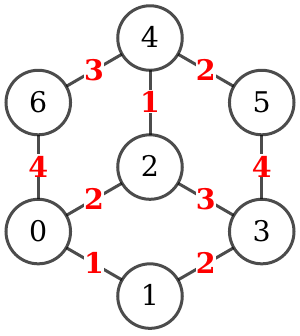}
    \includegraphics[scale=0.35]{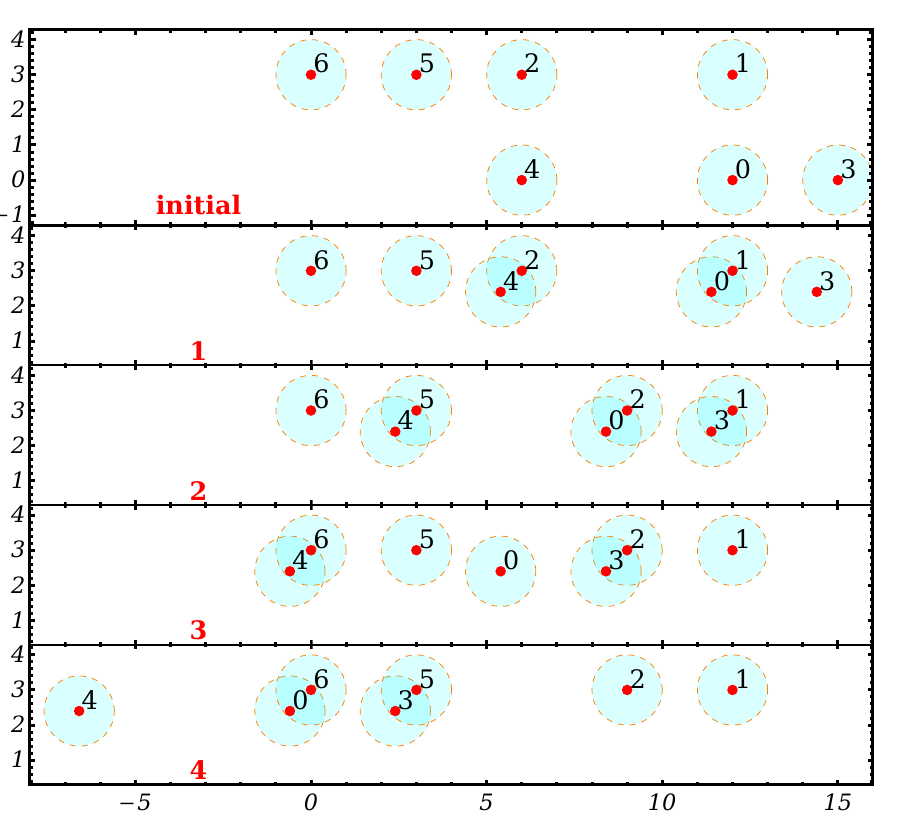}
    \label{fig:graphsteane}
    }
    \caption{ 
    Representation and preparation of a 12-qubit 1D cluster state and seven-qubit Steane code logical state.
    The graph representation of each graph state is shown using white circles and numbered edges;
    the quantum register configurations involved in the process are depicted in the tables.
    Numbered red dots represent atoms, blue disks indicate the blockade radii of the
    atoms, and numbered edges correspond to the atom configurations required to
    implement the entangling gates that operate concurrently. The initial register
    configuration is labelled ``initial,'' followed by a sequence of configurations
    implementing controlled-$Z$ gates. Having the atoms within the overlapping blockade radii, denoted by
    overlapping blue disks, is necessary to implement two-qubit gates between them.
    }
    \label{fig:graphs}
\end{figure}

Here, we replicate a recent neutral atom experiment presented
in~\cite{bluvstein2022quantum} in our virtual neutral atom device. The experiment involves the preparation of a
1D linear cluster state and a seven-qubit Steane code logical state, followed
by stabiliser measurements to benchmark the resulting states. Both of these
states are graph states~\cite{hein2004multiparty}, which can be efficiently
represented using graphs as illustrated in \Cref{fig:graphs}. In these graphs,
vertices correspond to plus states $\ket+\coloneqq(\ket0+\ket1)/\sqrt{2}$, and
edges represent controlled-$Z$ gates.  The preparation of such states requires the generation
of multiple non-local connections, which can be carried out concurrently on a
neutral atom platform.

Given a graph $G=(V,E)$ that comprises a list of vertices $V$ and edges $E$, the
corresponding graph state $\ket{\Psi_G}$ can be written as 
\begin{equation}
    \ket{\Psi_G}=\prod_{\{i,j\}\in E}\cz_{i,j}\bigotimes_{v\in V}\ket+_v 
\end{equation}
that is stabilised by a set of stabilisers
\begin{equation}
    \label{eq:stab}
    \mathcal S_{G}=\{X_j\bigotimes_{k\in N_G(j)} Z_k \mid j\in V\},
\end{equation}
where $N_G(j)$ is neighborhood of atom $j$. In the following,
we emulate the preparation of two graph states and measure their stabilisers
given graphs shown in \Cref{fig:graphs}.

\begin{figure}[hptb]
    \subfloat[]{\label{fig:stabgs}
    \includegraphics[scale=0.55]{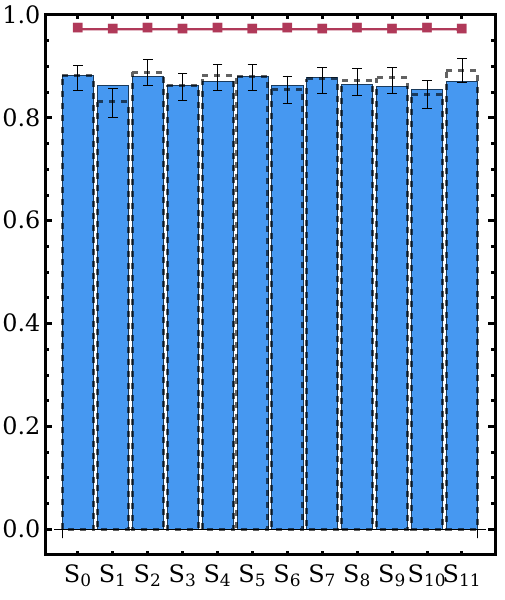}
    }
    \subfloat[]{\label{fig:stabsteane}
    \includegraphics[scale=0.55]{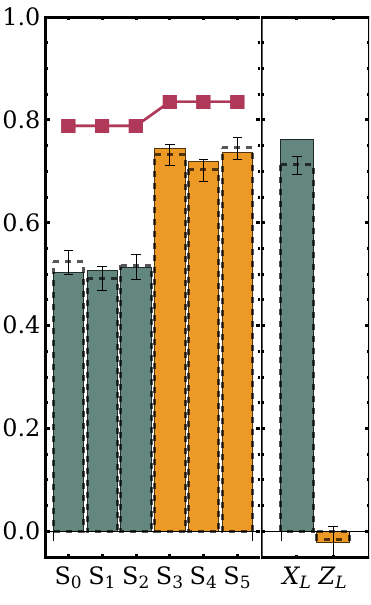}
    }
    \caption{Measurement statistics of:
        (a) stabilisers of the 1D cluster state with the graph shown in \Cref{fig:graph1d}, and
        (b) stabilisers (left) and logical $X$ and $Z$ measurements (right) of the seven-qubit Steane code prepared in the logical state $\ket+$, with the graph shown in \Cref{fig:graphsteane}. The green and orange bars indicate plaquette operators of $X$ ($S_0,S_1$, and $S_3$) and $Z$ ($S_3,S_4,$ and $S_5$), respectively. We took 2000 shots 
        of simulated measurements for each operator. The square markers
        indicate the results without SPAM errors. The dashed bars denote the results from 
        the experiments performed in \cite{bluvstein2022quantum}. The error parameters
        used in this simulation are tabulated in \Cref{conf:neutral_atoms} in the Appendix.
    }
\end{figure}

The preparation of a 12-qubit linear 1D cluster state is achieved using the three sequential atomic configurations illustrated in \Cref{fig:graph1d}. 
We measure the expected values of each stabiliser $S_j$, where $S_j$ is defined by \Cref{eq:stab} with the graph given in \Cref{fig:graph1d}. 
The resulting expected values in \Cref{fig:stabgs} are obtained from 2000 simulated measurements. Error parameters are manually adjusted to fit the experimental results. The average of stabiliser measurements in the simulation is $\langle S_i\rangle=0.87$, which is equal to that in the experiment. Without the SPAM errors, we obtain $\langle S_i\rangle=0.97$, while the experiment yields $\langle S_i\rangle=0.91$.

Preparing a logical seven-qubit Steane code can be achieved using five register
configurations, as illustrated in \Cref{fig:graphsteane}. Here, the logical 
state $\ket+_L\equiv(\ket0_L+\ket1_L)/\sqrt{2} $ is prepared by applying the transversal Hadamard gate to the logical state $\ket0_L$, \ie applying a Hadamard gate to each qubit. The expected value of each stabiliser and logical Pauli operators $X_L$ and $Z_L$ are measured. The stabiliser is defined by \Cref{eq:stab} with the graph given in \Cref{fig:graphsteane}. The logical Pauli operators are defined as $X_L=X_0X_1X_3$ and $Z_L=Z_0Z_1Z_3$,
which ideally result in $\langle X_L\rangle=1$ and $\langle Z_L\rangle=0$ 
for the logical state $\ket+_L$. The measurement results are shown in \Cref{fig:stabsteane}, obtained by simulating 2000 measurements.
The average expected values of the plaquette $X$ and $Z$ operators are 
$\langle{S_X}\rangle=0.51$ and $\langle S_Z\rangle=0.73$, respectively,
which the numbers agree with the experiment.
The expected value of the logical Pauli measurements in the simulation
are $\langle X_L\rangle=0.76$ and $\langle Z_L\rangle=-0.02$, while the 
experiments yield $\langle X_L\rangle=0.71$ and $\langle Z_L\rangle=-0.02$.

It is important to emphasise that simplifications have been implemented for practical purposes. For example, as noted earlier we model the blockade as a hard-edged sphere, neglecting the finite range of dipole-dipole interaction -- in the magnitude of $R^{-6}$ -- that may impact the outcomes. 
Consequently, some discrepancies in the distribution profile and expected values compared to experimental results are to be anticipated.
However, in certain cases, we have
managed to achieve values that precisely match those obtained in the experiments.

\section{Silicon qubits}\label{sec:silicon_qubits}

\subsection{Silicon qubits physical system}

Electron spin qubits in semiconductor quantum
dots~\cite{loss1998quantum,vandersypen2019semiconductor} utilise the inherent
two-level property of the spin-$\frac{1}{2}$ of electrons to encode states
$\ket0$ and $\ket1$. This is achieved by confining individual electrons in an
electrostatic potential well -- a quantum dot -- that can be controlled with voltages while subjecting them to a magnetic field. A typical platform is composed of an array of quantum dots with one or three electrons in each
well~\cite{philips2022universal}. The dots containing three electrons are
typically utilised as a part of the spin-to-charge
readout method, \ie spin parity readout~\cite{seedhouse2021pauli}.  The
tunnelling barriers between quantum dots can be adjusted via dedicated gate
voltages to engineer electrically controlled two-qubit exchange interactions.
Furthermore, the fabrication of the platform is compatible with advanced
semiconductor manufacturing techniques in some cases. The same lithographic process that is used in commercial foundries to create silicon chips for devices such as laptops or smartphones, can be employed to create quantum-dot array devices.
Therefore, spin qubits possess the potential as a scalable quantum computing
platform due to their practicality in control, well-known fabrication
techniques, long coherence time, and the tiny size that may enable integration of many millions of qubits in a single chip.

There are several semiconducting materials proposed to implement a spin qubit
quantum processor, such as GaAs and germanium.  However, here, we focus on 
silicon as the material to host the spin qubits with the quantum dots being
defined in the ${}^{28}$Si quantum well of the ${}^{28}$Si/SiGe
heterostructure. In particular, our silicon VQD platform is configured to correspond to
a recent device at the University of Delft~\cite{philips2022universal}; our error model is configured by analysing the description and
characterisation of the device reported by the authors of that paper.

\subsection{Architecture and native operations on virtual silicon qubits}

\begin{figure}[hpbt]
    \centering
\begin{tikzpicture}
  \def\n{6}  
  \def\r{0.38}  
  \def\d{1.4}  
  \def\textsize{\small}  

  \draw[thick,fill=yellow] (\d+\r,\r/4) rectangle (\d+\d,-\r/4);
  \draw[thick,fill=yellow] (\n*\d-\d,\r/4) rectangle (\n*\d,-\r/4);

  \draw[->, line width=1pt, black, >=stealth, scale=0.5,dotted] (4*\d+\r,-0.4) -- (2*\d+2*\r,-0.4);
  \draw[->, line width=1pt, black, >=stealth, scale=0.5,dotted] (6*\d+\r,-0.4) -- (4*\d+2*\r,-0.4);

  \draw[->, line width=1pt, black, >=stealth, scale=0.5,dotted] (8*\d+2*\r,-0.4) -- (9*\d+2*\r,-0.4);
  \draw[->, line width=1pt, black, >=stealth, scale=0.5,dotted] (10*\d+2*\r,-0.4) -- (11*\d+2*\r,-0.4);

  \foreach \i in {1,...,\n} {
    \draw[thick,fill=white] (\i*\d,0) circle (\r);
    \node at (\i*\d,0) {\textsize $Q{\i}$};
    \pgfmathtruncatemacro{\j}{\i-1}
    \node[below] at (\i*\d, -0.4) {\j};
   }
  \foreach \i in {2,...,\n} {
    \draw[thick,dashed] (\i*\d-\d+\r,0) -- (\i*\d-\r,0);
    }

\end{tikzpicture}

\caption{
    Architecture of six spin qubits in silicon quantum
    dots.  Labels $Qj$ indicate the qubits aliases used in 
    reference~\cite{philips2022universal}, while the numbers below represent qubit
    indices used in our VQD's internal notation. The dashed lines indicate qubit connectivity
    for operating controlled-phase gates ($\cph{\theta}$).
    The dotted arrows indicate the direction of controlled-$X$ gates ($\crot$),
    with the arrow pointing to the target qubit.
    The yellow rectangles indicate dot pairs that support direct parity readout,
    where output ``1'' indicates even parity and output ``0'' indicates odd parity.
    Measurement and initialisation of the middle qubits ($Q_2,\dots,Q_5$) are done indirectly.
}
\label{fig:silicon} 
\end{figure}
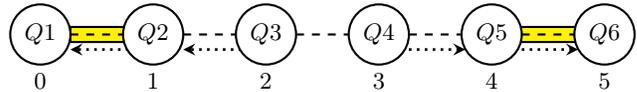

Our silicon virtual quantum device has a 1D linear architecture, with nearest-neighbour interactions. For simplicity, we take it that each quantum dot has one
electron, and tunnelling is allowed between dots to perform the projective
spin-to-charge conversion readout. We refer to the spin states down(up) as
$\ket{0(1)}$.  In fact, our VQD supports any even number of qubits; the user specifies the desired number to
determine the architecture (an even number of dots provides more practicality
in practice). For instance, \Cref{fig:silicon} shows the six spin qubits in
quantum dots used in our simulation.  

Direct readout of qubits is only possible for the pair of edge qubits, where
the obtained information is the parity of the two measured qubits.  The parity
measurement is realised via a projective measurement of the separated spin
configuration to the joint singlet subspace: states with even parity 00 and 11
are not allowed to tunnel whereas the $\ket{10}$ and $\ket{01}$ do, in
particular the $\ket{01}$ state via a fast decay. Moreover, the $\ket{01}$
decays (into $\ket{10}$) faster than the measurement timescale; such a protocol allows for a deterministic initialisation via real-time classical feedback. A
measurement operation can be simulated by a projective measurement $M$,
which projects a two-qubit state into the even and odd subspaces, followed by a
Kraus map $\mathcal K$ that simulates the decay of state $\ket{01}$.

The projective parity measurement $M$ is defined as 
\begin{equation}
\begin{gathered}
\label{eq:m}
    M =\{(\ketbra{01}+\ketbra{10},``0"),\\(\ketbra{00}+\ketbra{11},``1")\}, 
\end{gathered}
\end{equation}
where output ``0'' indicates detection of some 
change in current of a proximal detector, otherwise outputs ``1''. 
The Kraus map $\mathcal K$ comprises operators $K_j$, where 
\begin{equation}
    K=\{\ketbra{00},\ketbra{11},\ketbra{10},\ketbra{10}{01}\},
\end{equation}
and it fulfills $\sum_{j}K_j^\dagger K_j=I$. (Notice that the fourth element of $K$ differs from the others: it maps $\ket{01}$ to $\ket{10}$.) Thus, the parity
measurements, \eg on qubits $Q1,Q2$ from \Cref{fig:silicon} is simulated by the following
circuit
\begin{equation}
    \label{eq:projmeas}
    \begin{quantikz}[column sep=.2cm,row sep=0.1cm]
    \lstick{$\ket{\psi}_{Q1}$}& \gate[2]{\mathcal M}  \\
    \lstick{$\ket{\psi}_{Q2}$}&    & \\
    &     \texttt{p} \\
\end{quantikz}
\equiv
\begin{quantikz}[column sep=.2cm,row sep={0.8cm,between origins}]
\lstick{$\ket{\psi}_{Q1}$}& \qw     & \qw      & \ctrl{} & \gate[2]{\mathcal K}    & \qw\\
\lstick{$\ket{\psi}_{Q2}$}& \qw     & \ctrl{}  & \qw     &   & \qw \\
\lstick{$\ket{1}_a$}      &\gate{H} & \ctrl{-1}&\ctrl{-2}&\gate{H}&\meter{\texttt{p}}
\end{quantikz},
\end{equation}
where qubit $a$ indicates an ancillary qubit and $\mathtt p\in$\{``0'',``1''\} 
indicates parity measurement outcome of parity measurement projector $\mathcal M$ (\Cref{eq:m}).

To measure the middle qubits, we apply the controlled-$X$ ($\crot$) operation with the qubit of interest as controller, while employing the edge qubits as ancilla qubits; this is a form of quantum non-demolition (QND) measurement~\cite{yoneda2018quantum}. The following circuit achieves a QND measurement on $Q3$ and initialises qubits
$Q1,Q2$ into state $\ket{10}$ at simultaneously~\cite{philips2022universal}: 
\begin{equation}
    \label{eq:initqdots}
    \begin{quantikz}[column sep=.2cm,row sep=0.1cm]
        &\mathtt{p_1}&&\mathtt{p_2}&&\mathtt{p_3}\\
        \lstick{$\ket{\psi}_{Q1}$}& \gate[2]{\mathcal M} & \gate[cwires={1,1}]{X} & \gate[2]{\mathcal M}&\qw&\gate[2]{\mathcal M}&\qw \\
        \lstick{$\ket{\psi}_{Q2}$}&     & \qw &\qw &\targ{}&\qw & \qw \\[0.2cm]
        \lstick{$\ket{\psi}_{Q3}$}    & \qw & \qw &\qw &\ctrl{-1} &\qw & \qw\\,
\end{quantikz}
\end{equation}
where $\mathtt{p_3}$ is the outcome of measuring $Q3$ in the computational basis
and the flip gate $X$ is implemented conditioned on the classical bit $\mathtt{p_1}$.
By performing another flip $X$ to $Q3$ conditioned on the outcome
$\mathtt{p_3}$ followed by a repetition of the whole QND process, will result in deterministic initialisation of qubits 
$Q1,Q2$, and $Q3$ to state
$\ket{100}$. By applying the same procedures to qubits $Q4,Q5$, and $Q6$, utilising measurement from the other end of the chain,
we can achieve a deterministic full device initialisation into the state
$\ket{100001}$.


The elementary single-qubit quantum operations on this platform comprise
rotations $Rx(\theta)$ and $Ry(\theta)$, where $\theta$ is the rotation angle.
Single qubit rotations can be implemented on each qubit.  The rotation $Rz(\theta)$
can be implemented virtually (by adjusting the phase of the subsequent pulses).

The elementary two-qubit operations are controlled-phase gate
($\cph{\theta}_{i,j}$) and controlled-$X$ gate ($\crot_{i,j}$).
Controlled-phase gates can be implemented on any nearest neighbor qubits, as
illustrated by the dashed lines in \Cref{fig:silicon}. Controlled-$X$ gates
connectivity follows the dotted lines in \Cref{fig:silicon}, with the target
qubits indicated by the arrowhead.

\subsection{Error models for our silicon spin VQD}

The spin state readout $\mathcal M$ that is shown in \Cref{eq:projmeas} is the
native measurement operator in our virtual silicon platform.  Operator
$\mathcal M$ describes a two-qubit parity measurement accompanied with decay of
state $\ket{01}$ to $\ket{10}$. The error of this measurement is modelled by
two-qubit symmetric bit-flip error as described in \Cref{eq:bf2} that is
applied before and after applying the measurement operator $\mathcal M$.

Single qubit rotations are susceptible to two types of errors. The first is the
standard single-qubit error, for which error parameters are estimated using the
method discussed in \Cref{sec:one-qubit_gate}. In this device, we assume that
the error is dominated by the phase-flip error (\Cref{eq:deph1}). The second
type of error arises from the off-resonant Rabi oscillation, where the value
depends on the frequency detuning between the qubit that is operated upon and
the ESR frequency utilised to drive the qubit that characterises each qubit.

The two-qubit phase gates mediated by the exchange interaction,
such as $\cz$ or $\cph\theta$, also experience two
types of errors. Note that $\cz$ gate is the same type of gate as 
$\cph\theta$ that is achieved by setting $\theta=\pi$.
The first is the phase-flip error, as described in
\Cref{eq:deph2}. The second error is phase-flip cross-talk across the device
associated with residual off-state exchange coupling,
which is described by $Z$-rotation in \Cref{eq:weak_dephasing} and is dependent
on rotation parameter $\theta$. The strength of the cross-talk is characterised
experimentally and varies across different pairs of qubits. For the conditional
bit-flip ($\crot$), the error is modelled by two-qubit depolarising noise as
described in \Cref{eq:depol2}.

Passive qubits can experience free-induction exponential decays and weak
phase-flip cross-talk.  The exponential decays are characterised by the $T_1$
and $T_2$ time, which are described by \Cref{eq:depolt1} and \Cref{eq:depht2},
respectively. Weak dephasing cross-talk takes the form of conditional rotation,
as described in \Cref{eq:weak_dephasing} and is characterised experimentally.
This phase cross-talk is much stronger when operating a two-qubit phase gate
such as $\cz$ or $\cph\theta$. Here, we assume that the quantum gates are
implemented in a serial manner, where the effect of passive noise can be more
profound.

\subsection{Bell states preparation on silicon qubits}

In our virtual silicon qubits platform, we replicate a Bell pair generation
experiment using different pairs of qubits, as recently achieved 
in~\cite{philips2022universal}. We incorporate the reported experimental characteristics directly, and then the remaining parameters in the model are set through nonlinear optimisation based on the resulting Bell pair fidelities from the
experiment. Remarkably, we are able to obtain very similar fidelities and
output density matrices with a similar profile to the resulting tomography in
the experiment.

\begin{figure}[!t]\centering
\includegraphics[scale=0.5]{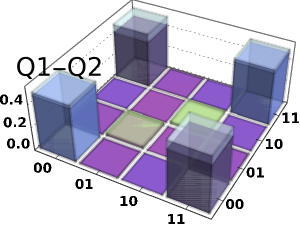}
\includegraphics[scale=0.5]{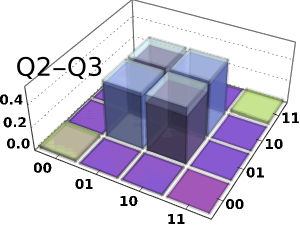}
\includegraphics[scale=0.5]{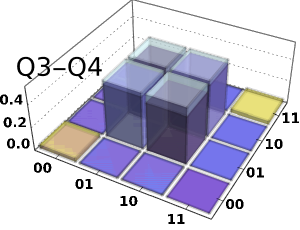}
\includegraphics[scale=0.5]{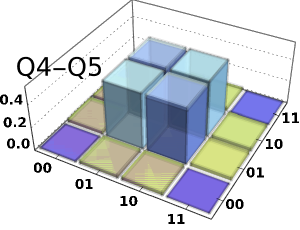}
\includegraphics[scale=0.5]{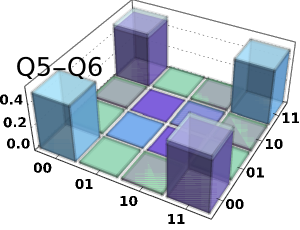}

Bell state fidelity(\%) and concurrency (\%) from 
the simulation and the experiment in~\cite{philips2022universal} (Exp) 

    \begin{tabular}{c|c|c||c|c}
      \toprule
Qubits&	Fid  &  Conc & Fid Exp & Conc Exp\\
      \hline
      Q1-Q2&	89.40&	79.75& 89.2$\pm$2.2&86.7$\pm$3.2\\
      Q2-Q3&	90.18&  80.42& 90.1$\pm$2.2&83.9$\pm$3.8\\
      Q3-Q4&	88.71&	79.08& 88.3$\pm$3.6&87.9$\pm$5.0\\
      Q4-Q5&	95.97&	94.49& 95.6$\pm$2.0&94.9$\pm$3.2\\
      Q5-Q6&	94.26&	90.59& 94.1$\pm$1.4&90.6$\pm$3.6\\
      \bottomrule
    \end{tabular}

    \vspace{1em}
Estimated fidelity (\%) of $\cz$ gates for each pair of qubits. Symbol $\mu$ indicates the average fidelity in the emulation and $\tilde\mu$ is the average fidelity in the experiment in~\cite{philips2022universal}.
\begin{adjustbox}{max width=0.5\textwidth}
    \begin{tabular}{c|c|c|c|c|c|c}
      \toprule
      \small Q1-Q2&
      \small Q2-Q3&
      \small Q3-Q4&
      \small Q4-Q5&
      \small Q5-Q6 &
      $\mu$ & $\tilde\mu$\\
      \hline
      93.7& 93.4& 92.9& 99.7& 97.9& 95.5 &92\\ 
      \bottomrule
    \end{tabular}
\end{adjustbox}
\caption{ Emulation of Bell pair generation experiment in
    \cite{philips2022universal} on the Silicon qubit virtual device. An
    optimisation is performed over the fidelity of $\cz$ (equivalently
    $\cph{\pi}$) gates subjected to the Bell pair generation fidelities in the
    experiment.  The optimised value of the $\cz$ fidelity is shown in the lower
    table. The estimated average fidelities of $\mathit{CZ}$ operations is 95.5, higher than
    the estimated value in the experiment, namely 92.  Using the optimised
    parameters, we plot the density matrix for each Bell pair overlapping with
    the noiseless case.
    The fidelity and concurrence of each Bell pair are shown in
the upper table. }\label{fig:Bells}

\end{figure}

The parameters we extract from the experiment detailed in \cite{philips2022universal}
are summarized in \Cref{conf:silicon} in the appendix. First, we incorporate their
experimental parameters and noise estimation into our simulation to the
greatest extent possible. Then, we optimize $\cz$ gate fidelity with a cost
function subject to the Bell state fidelities. We observe that $\cz$
fidelities have a dominant influence on the resulting Bell pair fidelities.
Finally, we obtain \Cref{fig:Bells}, which shows the density matrix of the resulting
Bell pairs and the optimized $\cz$ fidelities.

Our optimisation process is based on the fidelity of the Bell pairs, which has
allowed us to achieve a close match with the experimental results. Although
some resulting concurrencies are relatively poor, we have successfully obtained
a density matrix profile similar to the tomography result in the experiment for
most pairs. It is important to note that our density matrices are fetched
from the exact emulation, so we do not need to consider errors acquired during
the tomography process {including shot noise}. Overall, our results demonstrate that the VQD is a useful
tool to estimate processor parameters from experimental data and could be
utilised to guide the optimisation of Bell pair generation on a real device.

\section{Superconducting qubits}
\subsection{Superconducting qubits physical system}

Superconducting
qubits~\cite{devoret2004superconducting,kjaergaard2020superconducting} are
currently the most widely used and extensively studied solid-state qubits in
the quantum computing community. They have been the subject of numerous
advanced experiments, including the demonstration of quantum
advantage~\cite{arute2019quantum} and the implementation of NISQ (Noisy
Intermediate-Scale Quantum)
algorithms~\cite{riste2017demonstration,kandala2017hardware}.  

Our superconducting qubits VQD is based on superconducting transmon
qubits~\cite{transmon} that are connected by microwave resonators.  A transmon
qubit is a type of superconducting circuit consisting of a Josephson
junction~\cite{josephson1962possible} and a capacitance, while the resonators
provide a means for manipulating and reading out the qubits. In particular, our
model is based on a class of devices developed at the University of
Oxford~\cite{spring2022high,patterson2019calibration,rahamim2017double}.

In the circuit, the Josephson junction behaves as an oscillator, which can
exhibit either linear or non-linear behaviour depending on its operating regime.
When the junction is in its linear regime, it behaves like a simple harmonic
oscillator.  However, in the non-linear regime, the anharmonicity of the oscillator can be made sufficiently strong that we are able to regard the lowest two energy levels as a distinct two-level system: our qubit.
 Achieving high
non-linearity is critical to prevent leakage and improve the selective
manipulation and readout of the qubits.

\subsection{Superconducting qubits architecture and native operations}
\label{sec:architecture_scq}

\begin{figure}[hpbt]
\centering
\begin{tikzpicture}[circ/.style={draw, circle, inner sep=1pt}]
\draw (3,1.5) node[circ,label=above:{$Q_0$}] (q0) {4500};
\draw (3,0) node[circ,label=below:{$Q_1$}] (q1) {4900};
\draw (1.5,1.5) node[circ,label=above:{$Q_2$}] (q2) {4700};
\draw (1.5,0) node[circ,label=below:{$Q_3$}] (q3) {5100};
\draw (0,1.5) node[circ,label=above:{$Q_4$}] (q4) {4900};
\draw (0,0) node[circ,label=below:{$Q_5$}] (q5) {5300};
\draw[->, line width=1pt, black, >=stealth, scale=0.5] (q1) -- (q0) node [right, pos=0.3] {4};
\draw[->, line width=1pt, black, >=stealth, scale=0.5] (q3) -- (q2) node [right, pos=0.3] {4};
\draw[->, line width=1pt, black, >=stealth, scale=0.5] (q5) -- (q4) node [left, pos=0.3] {4};
\draw[->, line width=1pt, black, >=stealth, scale=0.5] (q2) -- (q0) node [above, pos=0.4] {1.5};
\draw[->, line width=1pt, black, >=stealth, scale=0.5] (q4) -- (q2) node [above, pos=0.4] {1.5};
\draw[->, line width=1pt, black, >=stealth, scale=0.5] (q5) -- (q3) node [below, pos=0.4] {1.5};
\draw[->, line width=1pt, black, >=stealth, scale=0.5] (q3) -- (q1) node [below, pos=0.4] {1.5};
\end{tikzpicture}

    \caption{\label{fig:sq2d}
    The layout of two-dimensional virtual six superconducting transmon qubits used in
    this simulation. 
    The nodes correspond to transmon qubits and edges indicate capacitive coupling between the qubits. The numbers displayed on the nodes represent the
    frequencies of the qubits, whereas those on the edges correspond to the coupling strength. All frequency values are expressed in MHz, and
    the edges are directed towards the higher qubit frequencies, which dictate the preferred  
    direction -- for a faster operation -- of the cross-resonance gate $\zx_{i,j}$, where qubit $Q_i$ has a higher
    frequency than qubit $Q_j$ ($Q_i\rightarrow Q_j$); in this case, we 
    call $Q_i$ as the control qubit and $Q_j$ as the target qubit.
    }
\end{figure}
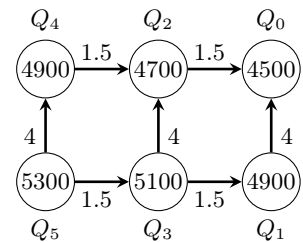

In our VQD, the end-user can provide the fundamental
characteristics of their superconducting qubits, which consist of qubit
frequencies, anharmonicities, and the exchange coupling strengths of the resonators.
The two-qubit `cross-resonance' gates~\cite{sheldon2016procedure,malekakhlagh2020first,patterson2019calibration} have a directional preference; this comes from its nature using an asymmetric microwave drive (driving the control qubit at the target qubit frequency), while the prefered direction is determined by the qubit frequency and anharmonicity configurations. As an example, the connectivity layout of the virtual
superconducting qubits in our simulation is illustrated in \Cref{fig:sq2d}.

Qubit initialisation must be executed at the beginning of the computation. Ideally, it sets all qubits to their ground state. In practice, most systems employ passive reset of superconducting circuits by waiting for an appropriate duration. During this process, the thermal bath from the surrounding environment is coupled with the qubits, allowing them to adiabatically reach thermal equilibrium, where some non-zero populations will reside in excited states.

Qubit readout corresponds to measurement in the computational basis. It is assumed that the measurement process accounts for signal amplification and classification processes~\cite{clerk2010introduction}.

The set of physical single-qubit gates comprises $Rx(\theta), Ry(\theta)$, and $Rz(\theta)$,
where $\theta\in[-\pi,\pi]$. The $Rz(\theta)$ gate is considered to be implemented by the virtual (noiseless) $Z$ rotation~\cite{mckay2017efficient}, which is accounted for in the phase of the subsequent pulse. 
In our model, the duration of physical single-qubit gates remains constant, irrespective of the variable $\theta$. This approach mirrors the common practice in real physical devices, where an arbitrary rotation gate is implemented by modulating the signal amplitude while maintaining a fixed gate duration.

The set of native two-qubit gates comprises the cross-resonance gates and the
siZZle (Stark-induced $\zz$ by level excursion)~\cite{wei2021quantum} gates:
\begin{align}
    \zx_{i,j} &\coloneqq e^{-i \frac{\pi}{2}Z_i\otimes X_j}\\
    \zz_{i,j} &\coloneqq e^{-i \frac{\pi}{2}Z_i\otimes Z_j},
\end{align}
respectively, which are usually implemented as non-parameterised gates.  
The siZZle gate is a two-qubit gate that relies on off-resonant driving to induce
an effective $\zz$-interaction between qubits. Both gates can 
be parameterised by adjusting the amplitude and the
driving pulse~\cite{kandala2017hardware}; however, fixing the parameters is more commonly practised. 
Therefore, both gates are non-parameterised in this model.

\subsection{Error models of superconducting qubits}

The qubit initialisation operation will set the qubits' state into a product state of
a statistical mixture of the ground state ($\ketbra0$) and excited state ($\ketbra1$), namely 
\begin{equation}\label{eq:thermalstate}
    \mathit{init}(\rho)
    =\bigotimes_{j}(p_j\ketbra0_j+(1-p_j)\ketbra1_j)
    \eqqcolon\rho_T,
\end{equation}
where $1-p_j$ is the probability associated with qubit $Q_j$ being excited from the ground state.
The resulting initialised state also corresponds to the thermal state $\rho_T$ at 
a finite small temperature; one may initialise the qubits by leaving them for a sufficient amount of time, although active routes are also used~\cite{cao2023emulating,gebauer2020state,magnard2018fast,riste2012feedback,riste2012initialization}.

Generalised amplitude damping $gamp(\Delta t)$  -- associated with a duration $\Delta t$ -- is used to model $T_1$-relaxation, which causes
the state to relax towards its thermal state $\rho_T$ (as described in
\Cref{eq:thermalstate}).  The action of $gamp(\Delta t)$ on a single-qubit density matrix $\rho$
is described as follows 
\begin{equation}
\begin{aligned}
    \label{eq:gdamp}
\rho& \mapsto \sum_{\alpha}\alpha \rho\alpha^\dagger,
\\
\alpha & \in
\Biggl\{
    \sqrt{p}\begin{pmatrix}
        1 & 0 \\
        0 & \sqrt{1-\gamma}
    \end{pmatrix},
    \sqrt{p}\begin{pmatrix}
        0 & \sqrt{\gamma} \\
        0 & 0
\end{pmatrix}, %
\\ &
\sqrt{1-p}
    \begin{pmatrix}
        \sqrt{1-\gamma} & 0 \\
        0 & 1
    \end{pmatrix}
,
 \sqrt{1-p}\begin{pmatrix}
        0 & 0\\
        \sqrt{\gamma} & 0
\end{pmatrix}
\Biggl\},\\
\text{and }\gamma&\equiv\gamma(\Delta t)=1-e^{-\Delta t/T_1},
\end{aligned}
\end{equation}
where $p$ is the probability of qubit population in state $\ketbra0$.
Therefore, for a very long duration, at $\Delta t\rightarrow\infty$,
the qubit will be in the statistical mixture of $p\ketbra 0+(1-p)\ketbra 1$.
On the other hand, the phase decoherence that is characterised by $T_2$-relaxation, is modeled with dephasing
noise, as described in \Cref{eq:depht2}.

The qubit readout error is characterised by the $T_1$-decay (\Cref{eq:gdamp}) and depolarising noise (\Cref{eq:depol1}) that is models as 
\begin{equation}
    \begin{quantikz}[column sep=0.2cm]
        \lstick{$Q_j$}& \gate{\mathit{gamp(t_M)}} & \gate{depol(\varepsilon)}& \meter{\texttt{0/1}} & \gate{\mathit{depol}(\varepsilon)}&\qw,
\end{quantikz}
\end{equation}
where $t_M$ indicates the measurement duration and $\varepsilon$ is infidelity
of the measurement. 
The $T_1$ decay is considered in order to accurately capture the decay occurring during the measurement process, which often takes a significantly longer duration (approximately 100 times longer than gate applications). The measurement is sandwiched between depolarising operators, accounting for misclassification error that comes from the noise from the received signal, where the first operator impacts the classical outcome, while the second one introduces uncertainty to the projected state.

In this model, the logical operators can be implemented perfectly and fully in
parallel. The errors arise solely from passive noise, which presents some
complexity. The passive noise comprises free induction exponential decays
characterised by $T_1$ and $T_2$, as well as the residual $\zz$-interactions from the static coupling along the capacitive coupling.

The capacitive coupling in a superconducting qubit system
can enable multi-qubit operations and entanglement, but they can also introduce
unwanted cross-talk. Specifically, in this model, we consider the static $\zz$
cross-talk that arises from an unintended or uncontrolled residual capacitive
coupling between two qubits~\cite{mundada2019suppression,mitchell2021hardware}.

If we denote the control qubit as $c$ and the target qubit as $t$, the
residual cross-talk is described as~\cite{krantz2019quantum}
\begin{equation}
    \exp(-i \beta Z\otimes Z), \quad \beta=J^2\left(\frac{1}{\Delta_{ct}-\alpha_t}-\frac{1}{\Delta_{ct}+\alpha_c}\right),
\end{equation}
where $J$ is capacitive coupling strength, $\Delta_{ct}$ is the frequency difference
between the qubits, and $\alpha_j$ is the anharmonicity of qubit $j$.
In this model, we assume that we 
can adjust the $\zz$ interaction strength to the desired value perfectly, incorporating the fact that a static $\zz$ crosstalk exists; thus, the static $\zz$ cross-talk is completely mitigated when implementing a siZZle gate.

\subsection{Variational Quantum Eigensolver on superconducting qubits}

The variational Quantum Eigensolver (VQE)~\cite{peruzzo2014variational} is an
promising route by which noisy-intermediate scale quantum (NISQ) devices may
solve practical problems. Notably, the algorithms have been demonstrated
to solve various small chemistry
problems~\cite{grimsley2019adaptive,ollitrault2020hardware,delgado2021variational,metcalf2020resource,chan2021molecular}.
Such algorithms are `hybrid', operating through a synergistic interplay between quantum and
classical computers, leveraging the strengths of both platforms. The achievable level of control on the quantum computer's part is of course a crucial aspect. Superconducting qubit platforms typically exhibit a variety of gates each with a degree of adjustability, \ie they are suitable as parameterised gates in VQEs. In the following, we
demonstrate a VQE for solving the ground state of $H_2$ molecule using the six-qubit superconducting VQD as described above and shown in \Cref{fig:sq2d}.  

The electronic Hamiltonian for the $H_2$ molecule is described using second quantisation, as follows:
\begin{equation}
    H = \sum_{i,j}h_{ij}a^\dagger_i a_j + \sum_{i,j,k,l}h_{ijkl}a^\dagger_ia^\dagger_j a_k a_l,   
    \label{eq:Hamiltonian}
\end{equation}
where $\{a_j\}$ and $\{a^\dagger_j\}$ are the lowering and raising operators that act
on a Fock space. Our choice of coefficients $h_{ij}$ and $h_{ijkl}$ represent the overlaps and exchange
integrals of the contracted three Gaussian functions of the STO-3G~\cite{szabo2012modern} basis set,
whose parameters are obtained via a self-consistent field procedure.
Then, the Fock space is mapped into the qubit Hilbert space using the canonical Jordan-Wigner
transformation~\cite{wigner1928paulische,seeley2012bravyi}. 
Finally, the Hamiltonian is expressed in the form of a Pauli sum,
\begin{equation}\label{eq:paulisum}
    H=\sum_k c_k P_k,
\end{equation}
where $c_k\in\mathbb R$ and ${P_k}\in\{I,\sigma_x,\sigma_y,\sigma_z\}^{\otimes
n}$ are $n$-qubit Pauli strings. We use the Python packages
\texttt{Openfermion}~\cite{mcclean2017openfermion} and
\texttt{PyScf}~\cite{sun2018pyscf} to generate the $H_2$ Hamiltonian that is
expressed as a Pauli sum spanning qubits $Q_0,Q_1,Q_2,$ and $Q_3$ of
\Cref{fig:sq2d}.

We use the VQE optimisation technique as described in
\cite{gustiani2022exploiting}.  The ansatz is constructed adaptively with the
gate pool comprising native superconducting qubit operators (see
\Cref{sec:architecture_scq}) and with connectivity shown in \Cref{fig:sq2d}.
The ansatz is iteratively refined, where in each iteration several intermediate
ans\"atze are given, then the one with the lowest energy is chosen and
simplified further by removing the superfluous gates.  We use the natural
gradient descent (also called imaginary-time evolution)
method~\cite{mcardle2019variational} to train our parameters. Finally, we
execute the VQE on our virtual superconducting qubits device, with error
parameters outlined in \Cref{conf:scq} in the appendix.  The optimisation
results are presented in \Cref{fig:vqe}.

\Cref{fig:vqe} shows that the output of the VQD deviates significantly from the ideal output because of the finite noise channels present. This is of course to be expected; generally, VQE studies have shown that precise (\eg within `chemical accuracy') results cannot be 
determined without incorporating a noise mitigation technique. Fortunately, however, such methods can make a dramatic improvement~\cite{kandala2017hardware}. It would be natural to extend the VQD study we present here by examining the efficacy of different mitigation methods, such as zero-noise extrapolation, quasi-probability, and so forth~\cite{cai2022quantum}. However, it is
interesting to note that deterministic errors are well-tolerated by VQEs since they represent a kind of systematic shift to parameters. 

\begin{figure}[!t]
    \includegraphics[width=\columnwidth]{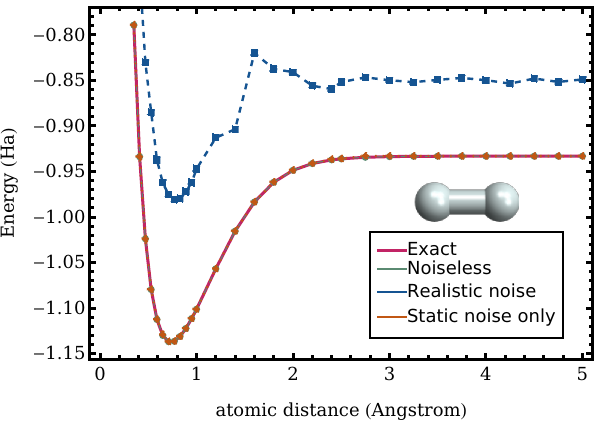}
    \includegraphics[width=\columnwidth]{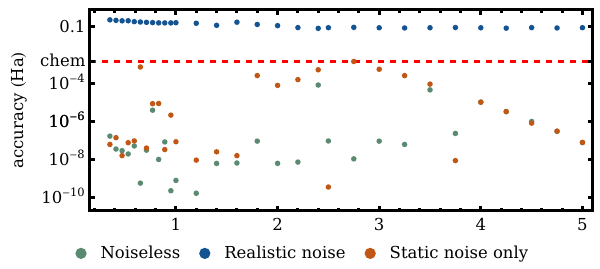}
    \caption{Estimations of ground state energies via VQE method (top) and the
        resulting accuracies (bottom), with the varying atomic distance of $H_2$
        molecule. The unit of energy is Hartree, with chemical accuracy shown
        in the red dashed line. The optimisation has been shown to be
        successful with the given quantum computer architecture, as indicated
        by the \emph{noiseless} line, which is aligned with the \emph{exact}
        line. Moreover, the optimisation is capable of handling unitary static
        noise, as indicated by \emph{static noise only} line. However, it fails
        when exposed to \emph{realistic noise} setting.}
    \label{fig:vqe}
\end{figure}

The six-qubit two-dimensional grid VQD described here does not correspond to any given reported experiment, but by making reasonable choices for the characteristics of the device we can narrowly predict its performance. This illustrates the utility of the VQD platform as a kind of fast-prototyping tool -- hardware teams can explore a range of systems and parameters in emulation in order to identify the most promising systems to actually fabricate.

\section{Conclusion and further discussion}

This paper has introduced the Virtual Quantum Device (VQD) tool and demonstrated its utility through a series of small studies -- each focusing on a different form of quantum computing technology. We have explored ion traps in the context of remote Bell state generation, NV-centre type architectures as simulators for BCS models, stabiliser measurements in neutral atom devices, Bell state preparation in silicon spin systems, and VQE algorithms with superconducting systems. In each case, we have linked to either state-of-the-art experiments (replicating the results of those papers with a close fit) or recently proposed theoretical ideas and protocols. 

These various demonstrations show that the VQD system is a valuable tool both for experimentalists wishing to understand data or test proposed designs, and for theorists who may want to explore ideas for algorithms and subroutines by testing them on a wide variety of platforms. Some of the studies are quite interesting investigations in their own right, and each of the VQDs that we created can straightforwardly be extended to match given physical devices even more closely. 

\smallskip

\section{Acknowledgements}

The authors would like to express their gratitude to the following contributors for their valuable contributions to this work:

Mohammed Alghadeer, Mohamed Abobeih, Alessandro Ciani, Shuxiang Cao, Andrew Daley, Joseph Goodwin, Fernando Gonzalez-Zalba, Tomas Kozlej, Peter Leek, David Nadlinger, Natalie Pearson, Gerard Pelergi, Cody Poole, Jonathan Pritchard, Mark Saffman, Jason Smith, and Joanna Zajac.

The authors would like to acknowledge the use of the University of Oxford Advanced Research Computing (ARC) facility~\cite{oxford_arc} in carrying out this work
and specifically the facilities made available from the EPSRC QCS Hub grant (agreement No. EP/T001062/1). CG is supported by this grant.
The authors also acknowledge support from EPSRC's Robust and Reliable Quantum Computing (RoaRQ) project (EP/W032635/1).

\bibliography{main} 


\appendix
\onecolumngrid

\newpage
\section{Design philosophy and choice of QuESTlink}
\label{appendix:designPhilosophyAndQL}
As noted in Section 2 of the main paper, there are a number of considerations that must inform the choice of which platform on which to base the VQD framework. We selected QuESTlink, because of its combination of the QuEST emulator (a fully featured and performant emulator) with the powerful symbolic manipulation capabilities of the commercial product Mathematica. Here we explain this choice and its benefits, both to expert users who will create new VQDs, and also for making our task of creating the VQD framework as tractable as possible. 

We noted that there are a number of potential pain points for the task of defining a new VQD.
    It requires describing the imperfect experimental processes, whether through canonical noise channels (like dephasing and depolarising) or more generally as Kraus maps, potentially requiring laborious tomography. Encoding the nuances of a real-time device into a dichotomy of active and passive, discrete noise processes can be a significant challenge for an experimentalist, and may require duplicated efforts to model unique devices of the same general platform (such as separate superconducting experiments).
    Further, it is often preferable to describe a device through a platform-native language (for instance, borrowing the language of superconducting experiments) and to model a device as a time-dependent object with its own internal states, rather than a static map between circuits and channels. 

        The virtual quantum device package must address these pain points. It provides a high-level and experimental platform-specific interface for generating device specifications, 
        where standard physical properties such as coherence times and coupling strengths become user-variable parameters. 
        In essence, the user configures a hardware platform in its native language and bespoke parameters, and the VQD generates a QuESTlink device specification in the language of digital gates and channels.

        These capabilities are achieved by making extensive use of Mathematica's functional, symbolic and pattern-matching facilities. 
        This significantly reduces the necessary code, and simplifies the software architecture, from a transpiler written in non-symbolic languages like {Python} or {C}. As a contrived example, consider the Mathematica snippet below:

    \begin{align*}
    &\texttt{u /. }\!\!\!\texttt{C}_{\texttt{c\_/;c<3}}\!\!
    \texttt{
    [(X|Y|Z)\!\!
    }_{\texttt{t\_/;t>=3}}
    \texttt{] /; \!\!\!\! (
        \!\!\!Abs[t-c] >= 2\!\!\!
    )}
    \\
    &
    \hspace{1.5cm}
    :> \texttt{Depol}_{\texttt{c,t}}\texttt{[15/16(1-Exp[-t])]}
    \end{align*}

    While it may appear opaque to users unfamiliar with Mathematica, this short sequence performs a rather complex task with only 71 keystrokes:
   It modifies the given QuESTlink circuit \texttt{u} (a list of symbolic gates) by replacing all controlled-Pauli gates which control upon the first $3$ qubits, and which target the remaining qubits excluding that adjacent to the control, with a two-qubit depolarising channel upon the same qubits with a target-dependent error probability. While this is an entirely artificial example, it showcases the expressibility of Mathematica and its utility for transpiling circuits.
 
         The VQD produces device specifications which can be subsequently passed to QuESTlink functions like \texttt{InsertCircuitNoise} in order to translate ideal circuits into realistic channels, expressed as a sequence of unitary and decohering operators. This sequence can then be exported, or effected upon a quantum state simulated with QuESTlink via functions like \texttt{ApplyCircuit}. 
        As noted in the main paper, all this applies both to state vectors and to density matricies according to the users preference, and this choice can be switched trivially from one to the other. 

\color{black}
\newpage
\section{Inferring the severity of depolarising and dephasing noise}\label{appendix:depolDephase}
Most reported gate fidelities in experiments come from random quantum state
benchmarking. Ideally, one measures the average fidelities over
Haar-distributed random quantum states. Using the results of
references~\cite{horodecki1999general,nielsen2002simple}, one can estimate the error
parameter for depolarising and dephasing channels (see
\Cref{sec:standard_forms}) given its average fidelity.  Note that in this
section, we only consider one- and two-qubit gates.

Let $\mathcal E$ a CPTP map describing errors of quantum gate $U$; both operators
are acting on Hilbert space $\mathcal H$, which has a dimension $d$. The 
\emph{average fidelity} $\bar F$ of map $\mathcal E$ is defined as 
\begin{equation}
    \bar F(\mathcal E)\equiv\int d\psi \bra{\psi}\mathcal E(\rho)\ket{\psi},
\end{equation}
where $\rho=\ketbra*\psi, \rho\in\mathcal H$, and the integral is normalised and evaluated over the
uniform (Haar) measure $d\psi$ on state space $\mathcal H$. Let the noisy form of 
the gate $U$ be $U_{\mathcal E}\equiv\mathcal E(U)$, then the
average fidelity of gate $U$ can be estimated by applying $U^\dagger$ to random states $\ket\psi$,
followed with $U_{\mathcal E}$,
\begin{equation}\label{eq:fbar}
    \bar F(\mathcal E)= \bar F(U_{\mathcal E},U)=\int d\psi \bra{\psi}U_{\mathcal E}( U\rho U^\dagger)\ket{\psi},
\end{equation}
where the perfect fidelity is obtained when $\mathcal E$ is an identity operation.
Second, the \emph{entanglement fidelity} of map $\mathcal E$ is defined as  
\begin{equation}\label{eq:def_entanglement_fidelity}
    F_e(\mathcal E)\equiv\bra{\Phi}(\mathcal E\otimes\mathcal I)(\rho_{\Phi})\ket{\Phi},
\end{equation}
where $\rho_{\Phi}\in\mathcal H\otimes\mathcal H, \rho_\Phi=\ketbra*{\Phi}$, and $\ket{\Phi}=\frac{1}{d}\sum_{j=0}^{d-1}\ket{jj}$ is a maximally entangled state, and $\mathcal I$ an identity operator acting on $\mathcal H$.

The entanglement fidelity of noisy gate $U_{\mathcal E}$ can be obtained as follows 
\begin{equation}\label{eq:fidelity_entanglement}
    F_e(\mathcal E)=F_e(U_{\mathcal E},U)=\bra{\Phi}(U_{\mathcal E}\otimes\mathcal I)(U\rho_{\phi}U^\dagger)\ket{\Phi},
\end{equation}
where $U$ is acting on the first half of the registers containing $\rho$.

Obtaining $F_e(\mathcal E)$ is more straightforward than $\bar F(\mathcal E)$.
Using \citeauthor{horodecki1999general}'s equation, we obtain the average
fidelity given the entanglement fidelity~\cite{horodecki1999general}: 
\begin{equation}\label{eq:fidelities_relation}
    \bar F(\mathcal E)=\frac{d F_e(\mathcal E) + 1}{d+1},
\end{equation}
where $d$ is the dimension of $\mathcal H$. Using this formula, we 
reverse-engineer the noisy gate given an average fidelity $\bar F$ that is
obtained by random state benchmarking in the experiment. We refer to
\cite{horodecki1999general,nielsen2002simple} for further details on this
fidelities relation.

\subsubsection{One-qubit gate error parameter estimate}\label{sec:one-qubit_gate}

Let $U_{\mathcal E}$ be a noisy single unitary matrix in which the error
$\mathcal E_{p,q}$ admits decomposition as in \Cref{eq:stderr}. First, we
calculate entanglement fidelity
\begin{equation}
    F_e=\bra*{\Phi^+}(\mathcal E_{p,q}\otimes\mathcal I)(\rho)\ket*{\Phi^+}, 
\end{equation}
where $\ket{\Phi^+}=(\ket{00}+\ket{11})/\sqrt{2}$ and $\rho=\ketbra*{\Phi^+}$.
Evaluating the value while keeping $p$ and $q$ as free (error) parameters gives us
\begin{align}
    F_e=1-p-q+\frac{4}{3} p q\label{eq:fe1}.
\end{align}

In practice, we want the capability to adjust the severity of depolarising and
dephasing errors. So now, we introduce $x\in[0,1]$ as a fraction describing depolarising
versus dephasing, where $x=1$ indicates completely depolarising, and $x=0$
signifies entirely dephasing.
Thus, we solve 
\begin{equation}
    1-p(1-x)-qx+\frac{4}{3} p qx(1-x)=\frac{3\bar F-1}{2}
\end{equation}
to estimate $p$ and $q$, where $\bar F$ is the reported one-qubit gate
fidelity in the experiment.

\subsubsection{Two-qubit gate error parameter estimate}\label{sec:two-qubit_gate}

If $U_{\mathcal E}$ is a two-qubit gate, we start with two pairs of Bell states
$\rho=\ketbra*{\Phi^+\Phi^+}_{p,r,q,s}$. Entanglement fidelity is defined as the
fidelity of the initial state after the application of the standard two-qubit noise $\mathcal E_{p,q}$,
\begin{equation}
    F_e=\bra*{\Phi^+\Phi^+}(\mathcal E_{p,q}\otimes\mathcal I)(\rho)\ket*{\Phi^+\Phi^+}_{p,r,q,s},
\end{equation}
where the error $\mathcal E_{p,q}$ is composed of two-qubit depolarising noise (\Cref{eq:depol2})
and two-qubit dephasing noise (\Cref{eq:deph2}). Then, we obtain the quantity
\begin{equation}
    F_e=1-p-q+\frac{16}{15}pq.
\end{equation}
Introducing error fraction (depolarising versus 
dephasing) $x\in[0,1]$ to the equation, we solve 
\begin{equation}
    1-p(1-x)-qx+\frac{16}{15}pqx(1-x)=\frac{5\bar F-1}{4}
\end{equation}
to estimate error parameters $p$ and $q$, where $\bar F$ is the reported
average two-qubit gate fidelity in the experiment.

\section{Error parameters used in simulations}

In this section, we provide a comprehensive overview of the parameters utilised within our virtual devices. This is intended to enable a precise reproduction of the simulations detailed within this study.

The following points offer a general overview of the parameters we employed. 
\begin{itemize}
    \item Time is expressed in the unit of microseconds ($\mu s$), frequency is in the unit of megahertz (MHz), and distance is expressed in micrometres $(\mu m)$, by default, unless stated otherwise; \eg see the case of NV-center.
    \item Gate fidelity of parameterised gates is assumed to be obtained via a randomised benchmarking method.
    \item When $T_2$ is used, we assume some Hahn-echo or dynamical-decoupling sequences are implicitly applied to or integrated into the subsequent pulses of the passive qubits. 
    \item   In practice, the severity of noise has a complex relation to the obtained gate, \eg $X$ vs $\sqrt{X}$. We approximate this by the parameter angle; namely, given the error parameter $\varepsilon$ and gate parameter $\theta$, the used error parameter is scaled to $\tilde\varepsilon\coloneqq\varepsilon\frac{\abs{\theta}}{\pi}$, where $\tilde\varepsilon$ should be in the proper range, \eg $\tilde\varepsilon\in[0,\frac{1}{2}]$ for single-qubit dephasing noise parameter. Such an approximation applies unless stated otherwise, \eg in the case of superconducting qubits gate duration is identical regardless of the gate parameter. 
\end{itemize}
\begin{table}[hp]
\caption{Variables used within the virtual two-node ion traps on the entanglement distillation. Here we assume every ion is identical on both nodes. }
\label{conf:trapped_ions}
    \begin{tabular}{p{3cm}p{3.5cm}p{9.3cm}}
\toprule
\textbf{Variable}&\textbf{Value}&\textbf{Description}\\ 
\midrule
\texttt{Nodes}&
\{(Alice,4), (Bob,4)\}&
The number of ions on each specified node.
\\ 
\texttt{T1}&
$3\times10^9$&
$T_1$ value of each qubit.
\\ 
\texttt{T2s}& 
$10\times10^5$&
$T_2^*$ value of each qubit.
\\ 
\texttt{StdPassiveNoise}&
\texttt{True}&
Apply the standard free-induction $T_1$ and $T_2^*$ decay on passive qubits.
\\ 
\texttt{DurMove}&
\{shuttling:25, split:50, combine:50, swap location:10\}&
Duration for each moves.
\\ 
\texttt{FidInit}&
0.9999&
Fidelity of the simultaneous ions initialisation in a zone.
\\ 
\texttt{DurInit}&
20&
Duration of the simultaneous ions initialisation in a zone.
\\ 
\texttt{DurRead}&
50&
Duration of a single ion readout.
\\ 
\texttt{BFProb}&
0.001&
Probability of bit-flip error in the readout process.
\\ 
\texttt{FidSingleXY}&
0.99999&
Gate fidelity of the single-qubit $x$- and $y$- rotations.
\\ 
\texttt{EFSingleXY}& 
1:0&
Error ratio \emph{depolarising:dephasing} when applying single-qubit $x$ and $y$ rotations.
\\
\texttt{RabiFreq}& 
10 &
The average Rabi frequency of single-qubit gates.
\\ 
\texttt{FreqCZ}
& 0.1 
&
Gate frequency in implementing $\cz$ gate.
\\ 
\texttt{FidCZ}& 
0.999 &
Gate fidelity of $\cz$ operator.
\\ 
\texttt{EFCZ}&
0.1:0.9&
Error ratio \emph{depolarising:dephasing} when applying $\cz$ gate.
\\ 
\texttt{FreqEnt}&
0.1&
Frequency of successful remote Bell pair generation.
\\ 
\texttt{FidEnt}&
0.95&
Fidelity of the generated remote Bell pair.
\\ 
\texttt{EFEnt}&
0.1:0.9&
Error ratio \emph{depolarising:dephasing} when obtaining the remote Bell pair.
\\
\bottomrule
\end{tabular}

\end{table}
\begin{table}[hp]
\caption{Variables used in the virtual NV-center diamond qubits on the simulating the dynamics of a BCS model. The following setting corresponds to the NV-center with five qubits $\{q_0,\dots,q_4\}$, where $q_0$ indicates the NV${}^-$ electron spin and the other $\{q_1,\dots,q_4\}$ correspond to the nuclear ${}^{13}$C spin. Time is expressed in seconds ($s$) and frequency is expressed in Hertz (Hz).}
\label{conf:nvc}
\begin{tabular}{p{2.5cm}p{4.1cm}p{9cm}}
\toprule
\textbf{Variable}&\textbf{Value}&\textbf{Description}\\ 
\midrule
\texttt{QubitNum}&
5&
The number of physical active qubits for computations.\\ 
\texttt{T1}&
\{3600, 60, 60, 60, 60\}&
$T_1$ values of qubits $\{q_0,\dots,q_4\}$, respectively.
\\
\texttt{T2}&
\{1.5, 10, 10, 10, 9\}&
$T_2$ values of qubits $\{q_0,\dots,q_4\}$, respectively.
\\
\texttt{FreqWeakZZ}&
3&
The frequency of weak $\zz$-coupling among nuclear spin qubits, on passive qubits.\\ 
\texttt{FreqSingleXY}&
\{15000, 0.5, 0.5, 0.5, 0.5\}$\times 10^3$
& 
Rabi frequencies of the single-qubit $x$- and $y$- rotations on qubits $\{q_0,\dots,q_4\}$, respectively.  
\\ 
\texttt{FreqSingleZ}&
\{32, 0.4, 0.4, 0.4, 0.4\}$\times 10^6$& 
Rabi frequencies for the single $z$-rotation on qubits $\{q_0,\dots,q_4\}$, respectively.\\ 
\texttt{FreqCRot}&
\{1.5, 2.8, 0.8, 2\}$\times 10^3$ &
Frequency of conditional rotations $\mathit{CRx}^\pm(\theta)$ and $\mathit{CRy}^\pm(\theta)$ on nuclear spins $\{q_1,\dots,q_4\}$, respectively, conditioned on the electron spin state.
\\ 
\texttt{FidCRot}&
\{0.98, 0.98, 0.98, 0.98\}&
Fidelity of conditional rotations $\mathit{CRx}^\pm(\theta)$ and $\mathit{CRy}^\pm(\theta)$ on nuclear spins $\{q_1,\dots,q_4\}$, respectively, conditioned on the electron spin state. 
\\
\texttt{FidSingleXY}&
\{0.9995, 0.995, 0.995, 0.99, 0.99\}&
Gate fidelity of single $Rx$ and $Ry$ rotations on qubits $\{q_0,\dots,q_4\}$, respectively.
\\
\texttt{FidSingleZ}&
\{1, 1, 1, 1, 1\}&Fidelity of single $Rz$ rotation on each qubit obtained by random benchmarking.\\ 
\texttt{EFSingleXY}&
{0.75:0.25}&
Error ratio \emph{depolarising:dephasing} when applying single-qubit $x$ and $y$ rotations.
\\
\texttt{EFCRot}&
{0.9:0.1}&
Error ratio \emph{depolarising:dephasing} when applying conditional rotations $\mathit{CRx}^\pm$ and $\mathit{CRy}^\pm$.
\\
\texttt{FidInit}&
0.999&
Fidelity of direct initialisation, namely initialising electron spin qubit $q_0$.
\\ 
\texttt{DurInit}&
$2\times10^{-3}$&
Duration of direct qubit initialisation on the electron spin qubit $q_0$.
\\ 
\texttt{FidMeas}&
0.946 &
Fidelity of direct qubit measurement on the electron spin qubit $q_0$.
\\ 
\texttt{DurMeas}&
$2\times10^{-5}$&
Duration of direct qubit measurement on the electron spin qubit $q_0$.
\\ 
\bottomrule
\end{tabular}

\end{table}
\begin{table}[hp]
\caption{Variables used within the virtual (Rydberg) neutral atoms to simulate stabiliser measurements on a 1D cluster and a logical Steane code states. We assume every atom has identical characteristics.}
\label{conf:neutral_atoms}
\begin{tabular}{p{2.6cm}p{4cm}p{9cm}}
\toprule \textbf{Variable}&\textbf{Value}&\textbf{Description}\\ 
\midrule
\texttt{T2}&
$100\times10^6$&
The $T_2$ value of each atom.
\\
\texttt{VacLifeTime}&
$100\times10^6$&
Vacuum life time of each tweezer, before the atom lost to the environment.
\\
\texttt{RabiFreq}&
0.1&
Rabi frequency for single-qubit operators.
\\
\texttt{UnitLattice}&
3 &
The unit length used in atom coordinates. This also a key to access internal parameter of the virtual device.
\\ 
\texttt{BlockadeRadius}&
$1$&
Short-range dipole-dipole interaction of Rydberg atoms; this becomes the maximal distance requirement among atoms 
that implement multi-qubit gate operations.
\\ 
\texttt{ProbLeakInit}&
0.01&
Leakage probability in the atom initialisation, where some population is excited outside the computational basis.
\\
\texttt{DurInit}&
$5\times10^5$&
Duration of qubit initialisation that happens in the beginning. This involves atom loading and characterising.
\\ 
\texttt{FidMeas}&
0.975&
Fidelity of qubit measurement.
\\ 
\texttt{DurMeas}&
10&
Duration of qubit readout.
\\
\texttt{ProbLossMeas}&
0.0001&
Probability of atom loss out of the optical trap during measurement process.
\\ 
\hline
\multicolumn{3}{c}{\textbf{1D cluster state simulation}} 
\\
\hline 
\texttt{QubitNum}&
12
&The number of atoms or qubits for computations.
\\ 
\texttt{AtomLocations}&
\{(0,0), (1,0), (2,0), (3,0), (4,0), (5,0), (6,0), (7,0), (8,0), (9,0), (10,0), (11,0)\}
&
Initial (2D) coordinate of physical locations of atoms $\{q_0,\dots,q_{11}\}$, accordingly.
\\ 
\texttt{ProbLeakCZ}& 
$\alpha=0.02$, $\beta=0.0001$
& 
Leakage probability in executing (multi) controlled-$Z$; these values correspond to
the parameter in the CPTN map given in \Cref{eq:czleak}.
\\ 
\texttt{ProbBFRot}&
p(1$\mapsto$0)=0.001, p(1$\mapsto$0)=0.03
&
Probability of asymmetric bit-flip error during single rotation operation.
\\ 
\hline
\multicolumn{3}{c}{\textbf{Logical state of Steane code simulation}} 
\\
\hline 
\texttt{QubitNum}&
7
&The number of atoms or qubits for computations.
\\ 
\texttt{AtomLocations}&
\{(0, 1), (1, 1), (2, 1), (4, 1), (2, 0), (4, 0), (5, 0)\}
&
Initial (2D) coordinate of physical locations of atoms $\{q_0,\dots,q_{8}\}$, accordingly.
\\ 
\texttt{ProbLeakCZ}& 
$\alpha=0.11$, 
$\beta=0.0001$
&
Leakage probability in executing (multi) controlled-$Z$; these values correspond to
the parameter in the CPTN map given in \Cref{eq:czleak}.
\\ 
\texttt{ProbBFRot}&
p(1$\mapsto$0)=0.1, p(1$\mapsto$0)=0.0001
&
Probability of asymmetric bit-flip error during single rotation operation.
\\ 
\bottomrule
\end{tabular}

\end{table}
\begin{table}[hp]
\caption{Variables used within the virtual silicon qubits device on the simulation of Bell pairs generation.}
\label{conf:silicon}
\begin{tabular}{p{2.8cm}p{6cm}p{6.5cm}}
\toprule 
\textbf{Variable} & \textbf{Value} & \textbf{Description}\\
\midrule
\texttt{QubitNum} & 6 &The number of physical active qubits for computations.
\\
\texttt{T1}& $10^4$ & $T_1$ duration of each qubit.
\\
\texttt{T2}& \{14, 21.1, 40.1, 37.2, 44.7, 26.7\} & $T_2$ duration of qubits $\{Q1,\dots,Q6\}$, respectively.
\\
\texttt{QubitFreq}& \{15.62, 15.88, 16.3, 16.1, 15.9, 15.69\}$\times10^3$&The fundamental qubit frequency of qubits $\{Q1,\dots,Q6\}$
\\
\texttt{RabiFreq}& \{5, 5, 5, 5, 5, 5\}&The Rabi frequency of single-qubit $x$ and $y$ rotations for qubits $\{Q1,\dots,Q6\}$, respectively.
\\
\texttt{OffResonantRabi}& 
\texttt{True}& 
Apply off-resonant driving noise when applying single qubit rotations.
\\
\texttt{StdPassiveNoise}& 
\texttt{True} & 
Apply the standard free-induction $T_1$ and $T_2$ decay on passive qubits.
\\
\texttt{FidSingleXY}&
\{0.9977, 0.9987, 0.9996, 0.9988, 0.9991, 0.9989\}&
Fidelities of single-qubit $x$ and $y$ rotations when applied to qubits $\{Q1,\dots,Q6\}$, respectively. 
\\
\texttt{EFSingleXY}&
0:1 &
Error ratio \emph{depolarising:dephasing} when applying single-qubit $x$ and $y$ rotations.
\\
\texttt{FreqCZ}&
\{12.1, 11.1, 6.6, 9.8, 5.4\} &
Rabi frequency of controlled-$Z$ or -$Ph(\pi)$) gate when operated on qubit pairs $\{(Q1,Q2),\dots,(Q5,Q6)\}$, respectively.
\\
\texttt{FidCZ}&
\{{0.937, 0.934, 0.929, 0.997, 0.979}\}&
Fidelity of controlled-$Z$ or -$Ph(\pi)$ gate when operated on  qubit pairs $\{(Q1,Q2),\dots,(Q5,Q6)\}$, respectively. 
\\
\texttt{EFCZ}& 0:1 & Error ratio \emph{depolarising:dephasing} on the two-qubit gates error.
\\
\texttt{ExchangeRotOn}&
$\begin{pmatrix}0&0.023&0.018&0.03&0.04\\0.05&0&0.03&0.03&0.04\\0.05&0.03&0&0.07&0.042\\0.038&0.03&0.031&0&0.25\\0.033&0.03&0.02&0.03&0\end{pmatrix}$&
Cross-talk on the passive qubits in form of entangling rotation controlled-$Rz(\delta_{i,j}\frac{\theta}{\pi})$, where $\delta_{i,j}$ is the matrix element and $\theta$ is the angle of the gate; this applies when a two-qubit gate is active.
\\
\texttt{ExchangeRotOff}&
\{0.039, 0.015, 0.03, 0.02, 0.028\}&
Cross-talk on the passive qubits in form of entangling rotation controlled-$Rz(\alpha_{i,i+1})$, where $\alpha_{i,i+1}$ is specified here; this applies when no two-qubit gate is active.
\\
\texttt{FidRead}&
0.99&
Fidelity of the parity readout on the edges, between qubits $Q1,Q2$ and $Q5,Q6$. 
\\
\texttt{DurRead}&
10&
The total duration on the parity readout on the edge qubits: $Q1,Q2$ and $Q5,Q6$.
\\
\bottomrule
\end{tabular}

\end{table}
\begin{table}[hp]
\caption{Variables used within the virtual superconducting qubits to simulate a variational quantum eigensolver to approximate the ground state of a hydrogen molecule.}
\label{conf:scq}
\begin{tabular}{p{3cm}p{5cm}p{7.5cm}}
\toprule 
\textbf{Variable}&\textbf{Value}&\textbf{Description}\\ 
\midrule
\texttt{QubitNum}&
6&
The number of physical active qubits for computations.\\ 
\texttt{T1}&
\{63, 93, 109, 115, 68, 125\}&
$T_1$ values of qubits $\{Q_0,\dots,Q_5\}$, respectively.
\\
\texttt{T2}&
\{113, 149, 185, 161, 122, 200\}&
$T_2$ values of qubits $\{Q_0,\dots,Q_5\}$, respectively.
\\
\texttt{ExcitedInit}&
\{3.2, 2.1, 0.8, 0.9, 2.5, 0.7\}$\times10^{-2}$&
The probability of the excited population from the ground state ($p_j$) for qubits $\{Q_0,\dots,Q_5\}$,
respectively. The qubits will be initialised into statistical mixture of $\bigotimes_j(p_j\ketbra1+(1-p_j)\ketbra0)$.
\\
\texttt{QubitFreq}&
\{4.5, 4.9, 4.7, 5.1, 4.9, 5.3\}$\times 10^3$&
The fundamental qubit frequency of qubits $\{Q_0,\dots,Q_5\}$, respectively.
\\ 
\texttt{ExchangeCoupling}&
\{(0,1):4, (0,2):1.5, (1,3):1.5, (2,3):4, (2,4):1.5, (3,5):1.5, (4,5):4\}&
The capacitive coupling strength constants that allows exchange interaction between two transmons $(Q_i,Q_j)$ but it creates residual cross-talks on passive noise; 
thus, it has the form of $\zz$ interaction. The cross-talk can be mitigated by tuning the interactions accordingly, \ie when operating
a siZZle gate; thus, if a siZZle gate is on, the $\zz$ cross-talk is gone. 
\\ 
\texttt{Anharmonicity}&
\{296.7, 298.6, 297.4, 298.3, 297.2, 299.1\}&
The anharmonicity frequency of qubits $\{Q_0,\dots,Q_5\}$, respectively. In this model, it shifts the strength of the static $\zz$ cross-talk.
\\ 
\texttt{FidRead}&
\{0.9, 0.92, 0.96, 0.97, 0.93, 0.97\}&
Readout fidelity of qubits $\{Q_0,\dots,Q_5\}$, respectively.
\\ 
\texttt{DurMeas}&
5&
Duration of qubit readout, taking into account (classical) outputs classification process.
\\ 
\texttt{DurRxRy}&
0.05&
Gate duration to implement single-qubit rotations $Rx(\theta)$ and $Ry(\theta)$, that remains constant regardless $\theta$.
\\ 
\texttt{DurZX}&
0.5&
Gate duration to implement cross-resonance gates $\zx$. 
\\
\texttt{DurZZ}&
0.5&
Gate duration to implement siZZle gates
\\
\texttt{StdPassiveNoise}&
\texttt{True}&
Apply the standard passive noise, namely the free-induction $T_1$- and $T_2$-decays.
\\ 
\texttt{ZZPassiveNoise}&
\texttt{True}&
Apply the static $\zz$ interaction passive noise.\\\bottomrule
\end{tabular}

\end{table}

\newpage

\section{Entanglement distillation on trapped ions}\label{app:distillation}

The following figures show every sequence to perform three rounds of phase-flip error entanglement distillation. The left side shows the time and the operator performed at that time. The left side shows arrangement of the ions after applying the operator on the left side.

\begin{figure}[!h]
\caption{Operations and moves involved in three rounds of phase-flip distillation on the virtual trapped ions which comprises 112 sequences.}
\centering
\includegraphics[scale=0.53]{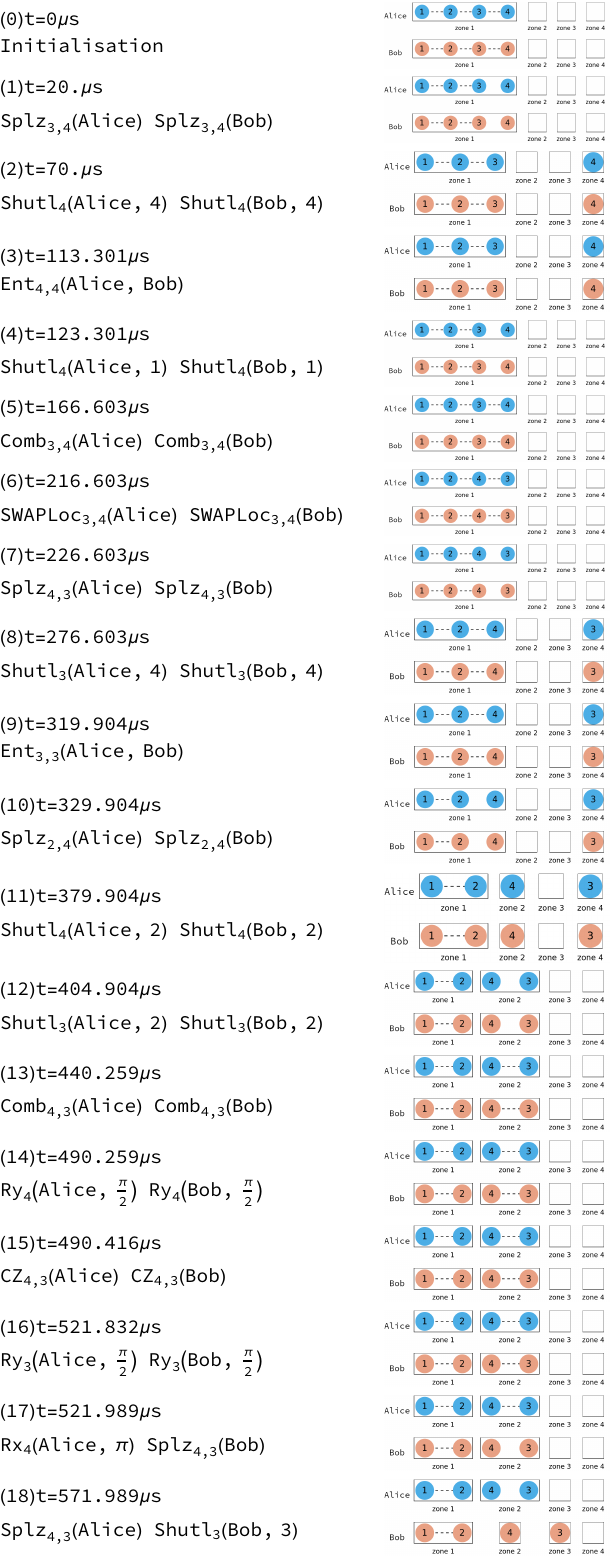}
\includegraphics[scale=0.53]{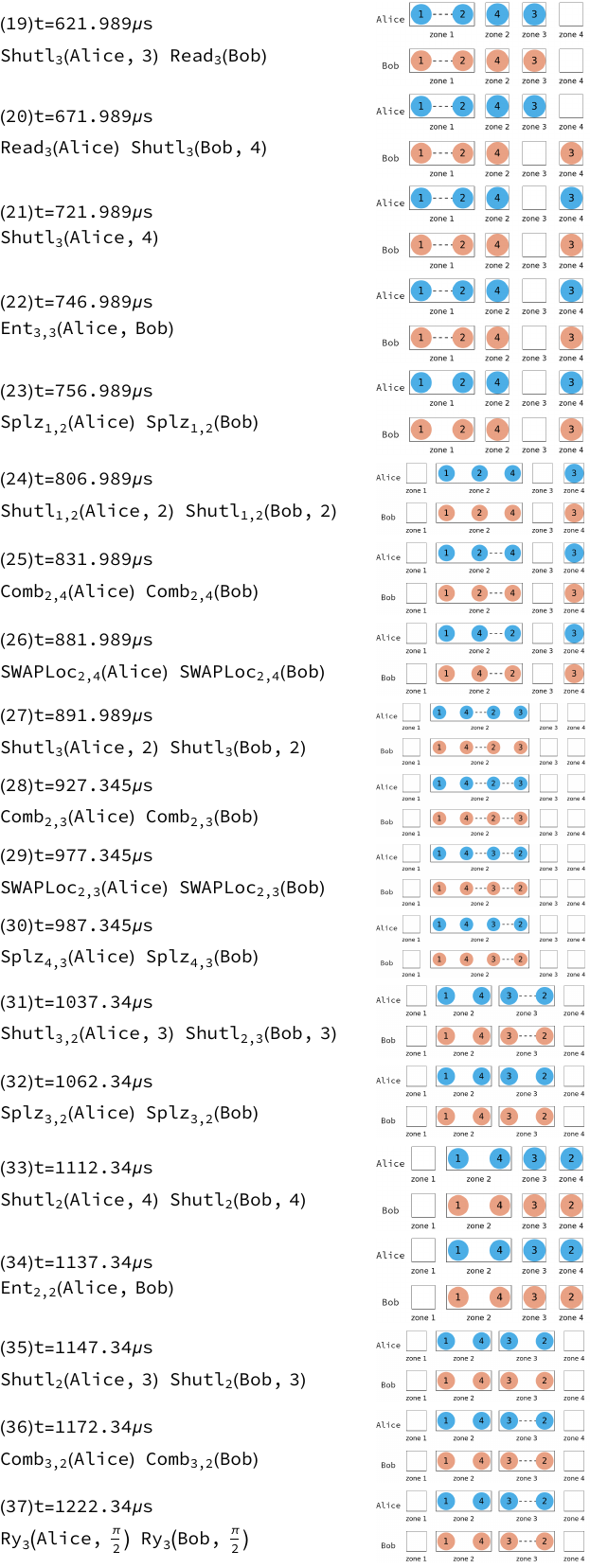}
\includegraphics[scale=0.51]{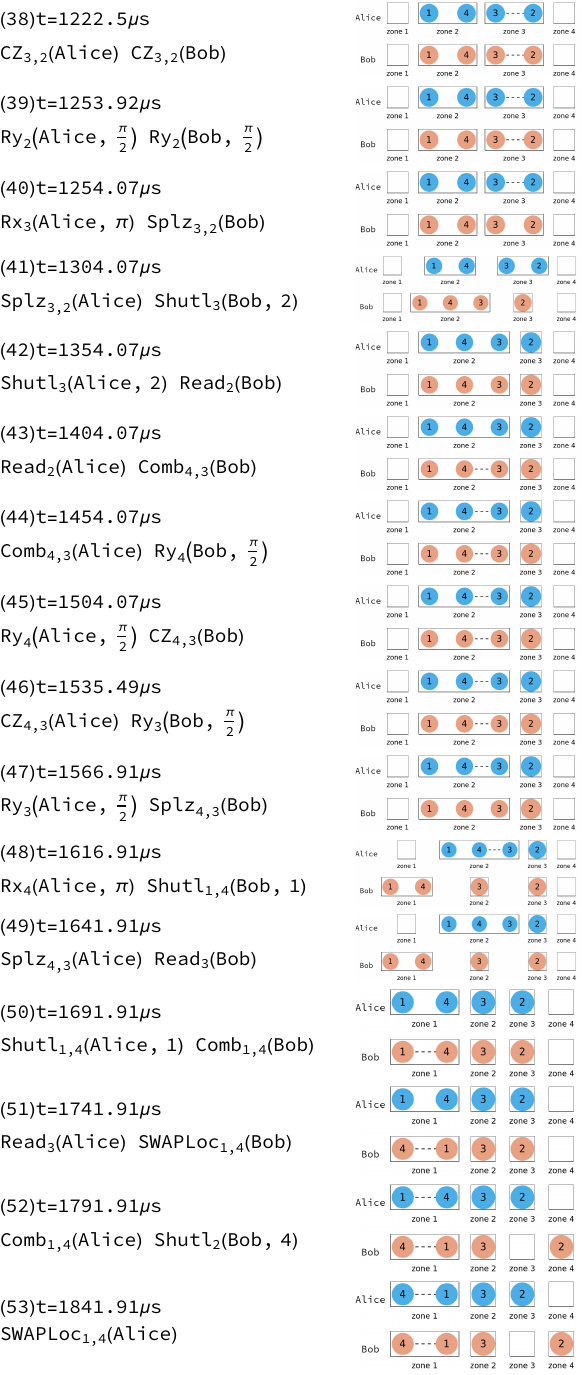}
\end{figure}
\begin{figure}[!h]
 \ContinuedFloat\centering
\includegraphics[scale=0.54]{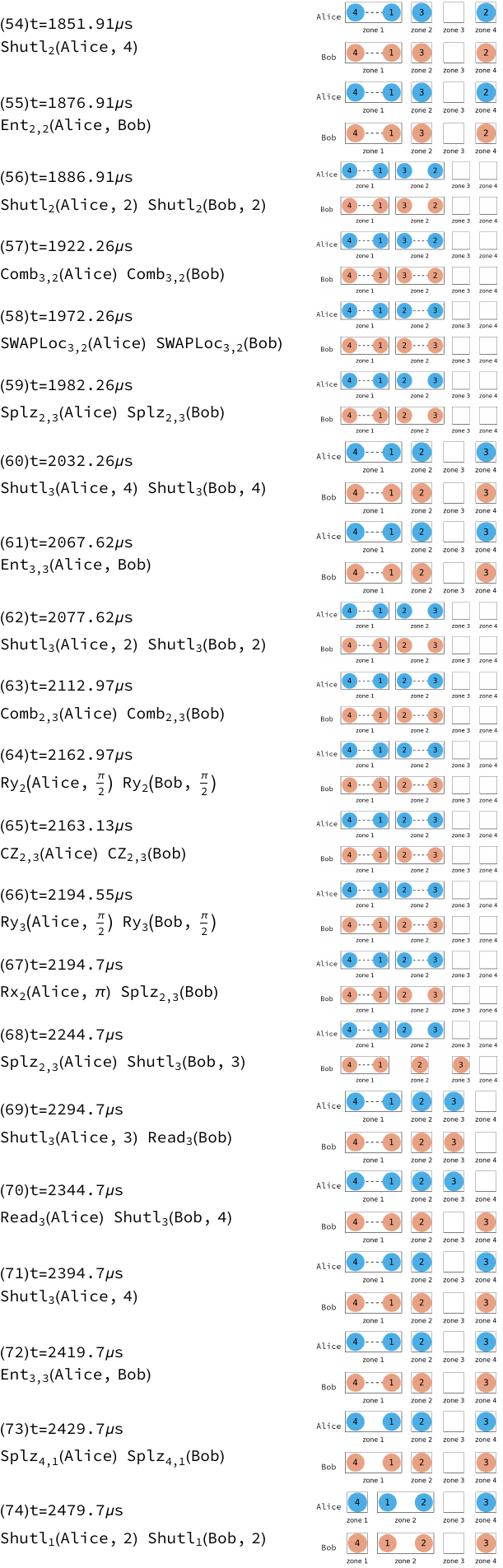}
\includegraphics[scale=0.52]{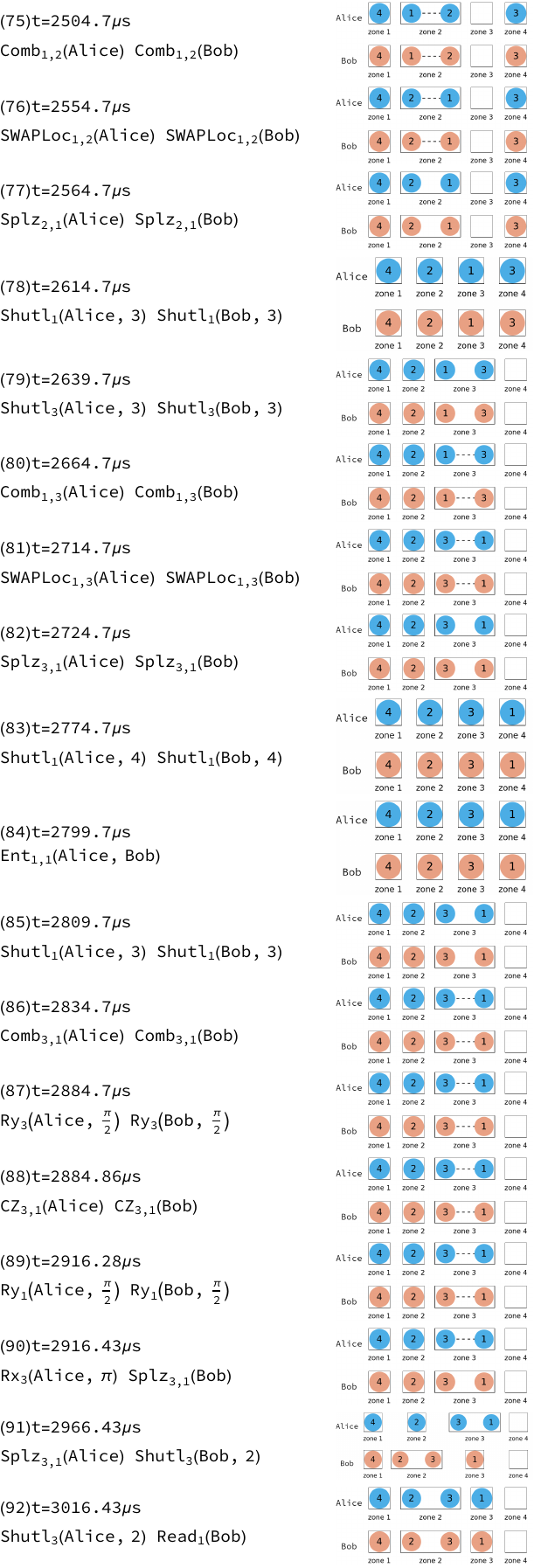}
\includegraphics[scale=0.51]{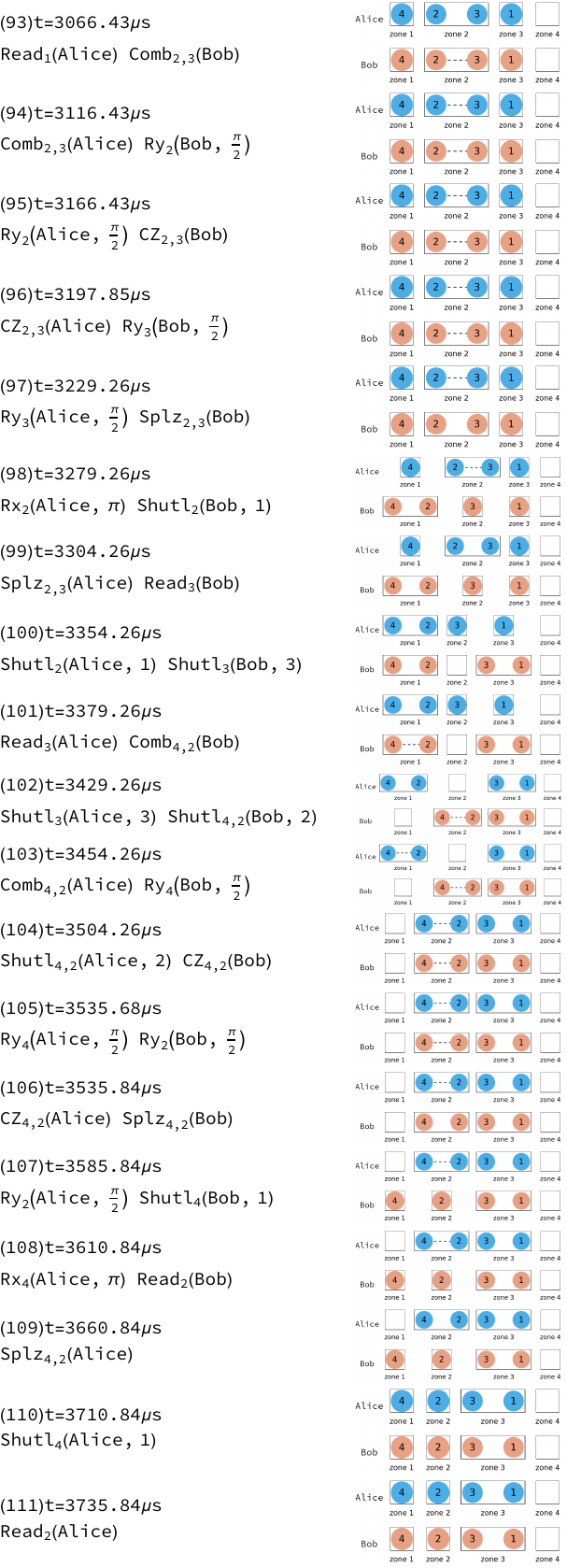}
\label{fig:distillation_steps}
\end{figure}

\newpage
\section{Parallelisation in Neutral Atoms}\label{app:parallel_na}

Parallelism of quantum gates in neutral atoms is subject to the Rydberg blockade mechanism. As illustrated in \Cref{fig:blockade}, gate parallelism is permitted for the qubits outside the restricted zone, \ie the resulting zone from the interacting atoms. Note that the end user is responsible for reconfiguring the register to accommodate such a blockade mechanism.

\begin{figure}[hbpt]
\includegraphics[scale=1.2]{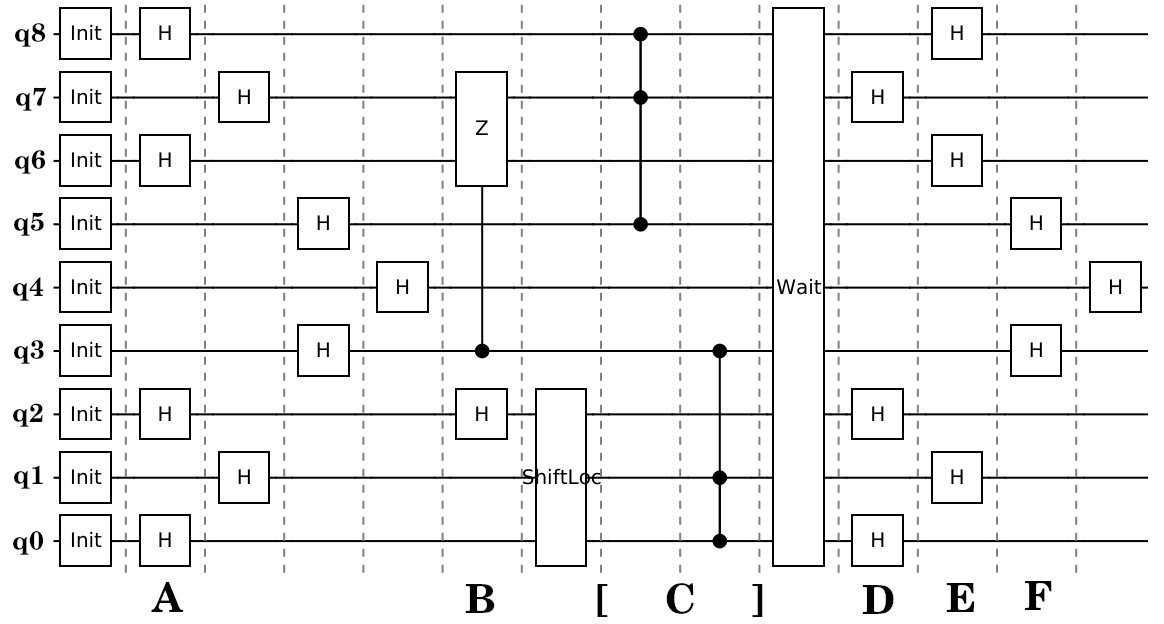}

\includegraphics[scale=0.35]{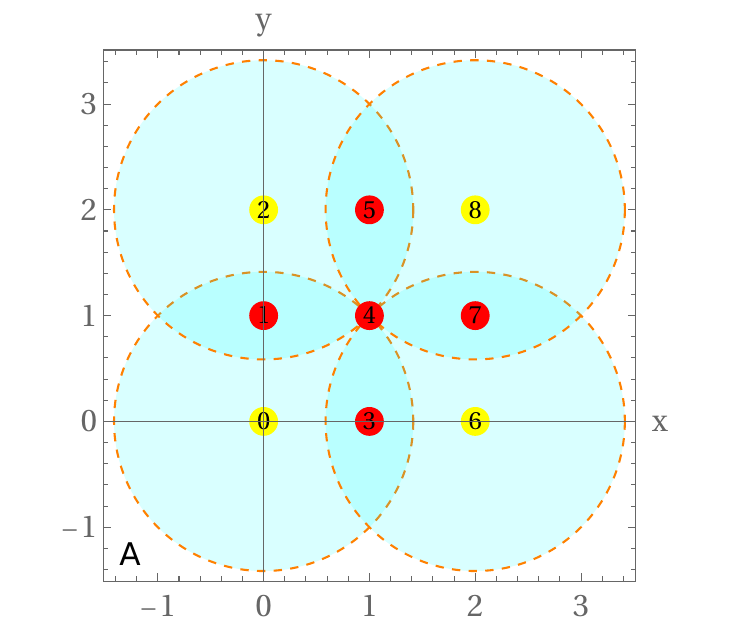}
\includegraphics[scale=0.35]{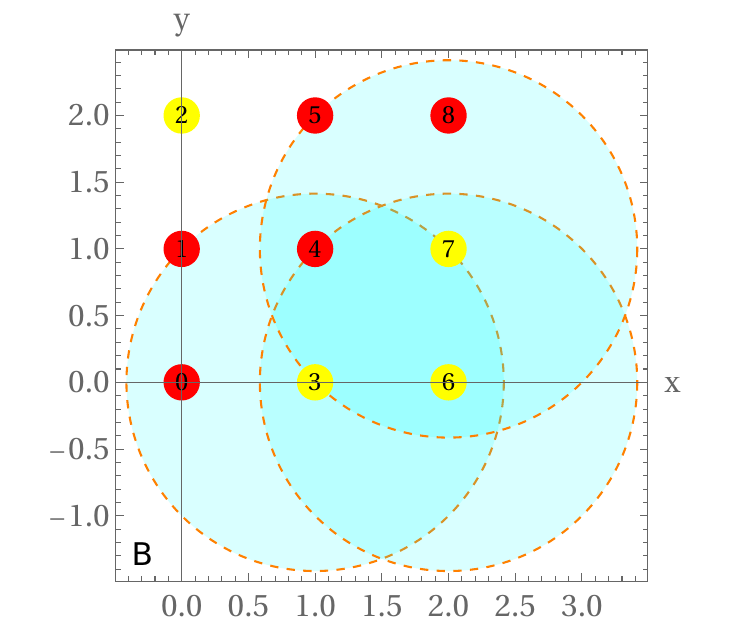}
\hspace{-1cm}
\includegraphics[scale=0.35]{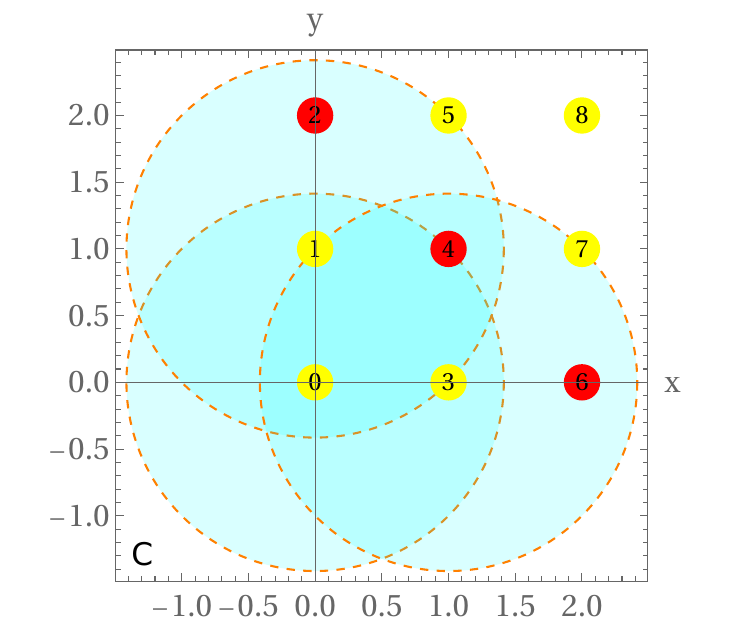}
\hspace{-.5cm}
\includegraphics[scale=0.3]{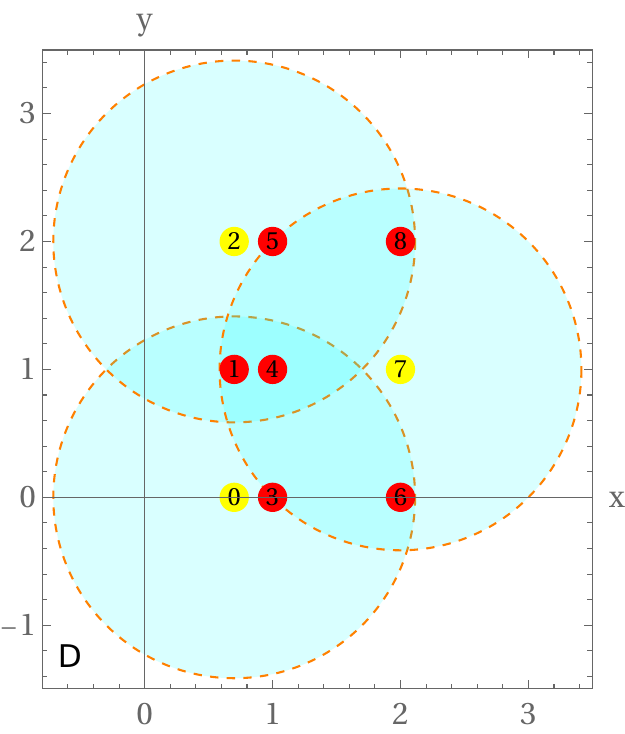}
\includegraphics[scale=0.3]{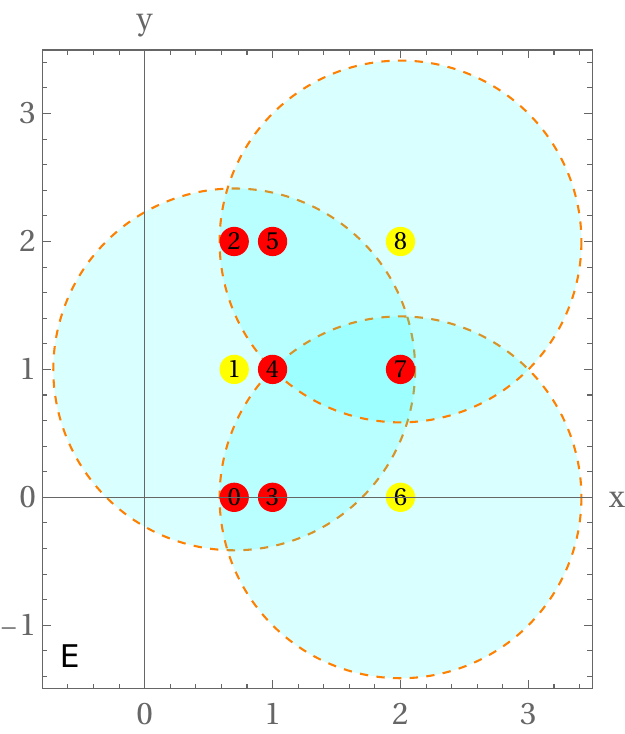}
\includegraphics[scale=0.3]{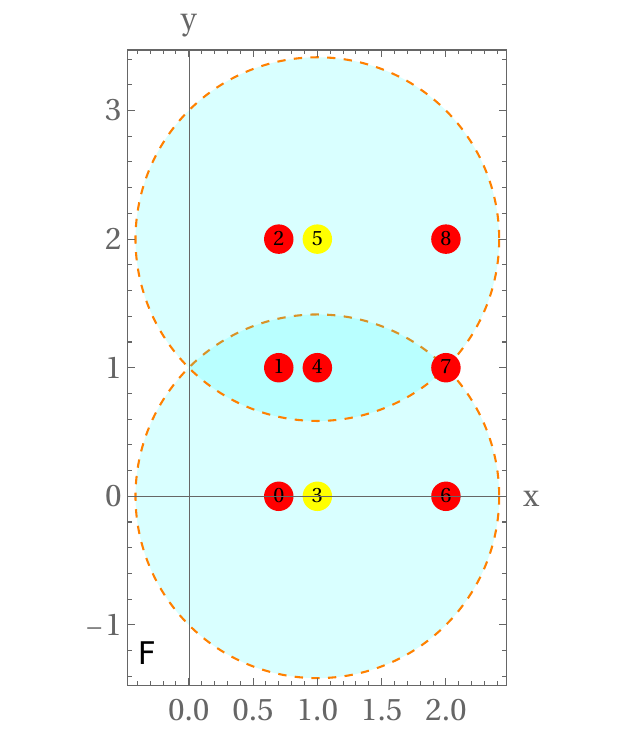}
\caption{\label{fig:blocakde_example}
In the circuit, qubit $q_j$ represents atom $j$ in panels \texttt{A}-\texttt{F}. The yellow atoms indicate active qubits, \ie a quantum operation is acted upon. All initialisation is done in parallel, which represents loading the atoms or preparing the register. At column \textbf{A}, Hadamard gates ${H_0, H_2, H_6, H_8}$ are applied simultaneously because, as Panel \texttt{A} illustrates, yellow atoms $0, 2, 6, 8$ are not within each other's interaction zones—unlike the red atoms. In column \textbf{B}, gate parallelization is feasible for a three-qubit gate acting on atoms $3, 6, 7$, along with a Hadamard $H_2$; this is viable because atom 2 is outside the restricted zone, as depicted in panel \texttt{B}. Here, we assume operations that reconfigure the register, such as \texttt{ShiftLoc} and \texttt{SWAPLoc}, are executed serially: no concurrent operations are permitted. Columns in \textbf{C} demonstrate that the two 3-qubit gates cannot be implemented simultaneously, as panel \texttt{C} indicates atoms 5 and 7 are within the restricted zone when the 3-qubit gate involving atoms 0, 1, 3 is activated. Following the reconfiguration of atoms as per panel \texttt{D}, the arrangement of Hadamard gates in column \textbf{D} differs from that in column \textbf{A}: the proximity of more atoms reduces the number of gates that can be parallelized. This principle also applies to columns \textbf{E} and \textbf{F}, corresponding to panels \texttt{E} and \texttt{F}, respectively.}
\end{figure}

In the following example, we have an array of 9 atoms in a square lattice with distance $d$, where each atom has blockade radii $r_b=\sqrt{2}d$. The first configuration is shown in \Cref{fig:blocakde_example} panels \texttt{A}-\texttt{C}, which is then reconfigured by shifting a column of atoms, as shown in \Cref{fig:blocakde_example} panels \texttt{D}-\texttt{F}. 

For a concrete example, We show the parallelism of a circuit comprises the following operations: 
\begin{enumerate}
    \item Initialise the quantum register: $\{\mathit{Init}_j\}$
    \item Apply Hadamard gates to all atoms: $\{H_j\}$
    \item Apply a three-qubit gate: $C_{3}[Z_{6,7}]$
    \item Shift location of atoms 0, 1, and 2 horizontally: $\texttt{ShiftLoc}_{0,1,2}[\{-x, 0\}]$
    \item Apply three-qubit gates: $\mathit{CZ}_{0,1,3}\mathit{CZ}_{5,7,8}$
    \item Do nothing: \texttt{Wait}
    \item Apply Hadamard gates to all atoms: $\{H_j\}$.
\end{enumerate}
The tool will schedule the circuit shown in \Cref{fig:blocakde_example} due to the blockade restrictions.

\end{document}